\documentclass{ws-rv961x669}
\usepackage{ws-rv-van}     
\usepackage{graphicx}
\usepackage{mathtools} 		
\usepackage{amssymb,amsmath,bbold}
\usepackage{float}	 
\usepackage{color}
\usepackage{hyperref}
\usepackage{doi}
\newcommand{\nc}{\newcommand} 
\newcommand{\rnc}{\renewcommand} 
\nc{\req}[1]{Eq.\,\ref{#1}} 
\nc{\rf}[1]{Fig.~\ref{#1}} 
\nc{\Th}{\ensuremath{T_\mathrm{H}\,}}
\nc{\RHI}{relativistic heavy ion}
\rnc{\topfraction}{.99} 
\rnc{\bottomfraction}{.99}
\rnc{\textfraction}{.0} 
\nc{\scs}{\scriptstyle}
\nc{\JP}{$J\!/\!\Psi$\,}
\nc{\lsim}{\,{\buildrel < \over {_\sim}}\,}
\nc{\gsim}{\,{\buildrel > \over {_\sim}}\,}
\nc{\ie}{{\it i.e.}}
\nc{\eg}{{\it e.g.}} 
\nc{\agev}{{\rm A\,GeV}\,}
\begin{document}

\chapter[QGP Discovery]{From Strangeness Enhancement to\\ Quark-Gluon Plasma Discovery\footnote{\phantom{1}}%
}
\author[Koch,  M\"uller, Rafelski]{Peter Koch$^1$, Berndt M\"uller$^2$ and Johann Rafelski$^3$
}
\address{$^1$Bitfabrik GmbH \& Co. KG, D-63110 Rodgau, Germany\\
$^2$Department of Physics, Duke University, Durham, NC 27708, USA\\
$^3$Department of Physics, The University of Arizona, Tucson, AZ 85721, USA }

\begin{abstract}
This is a short survey of signatures and characteristics of the quark-gluon plasma in the light of experimental results that have  been obtained over the past three decades. In particular, we present an in-depth discussion of the strangeness observable, including a chronology of the experimental effort to detect QGP at CERN-SPS, BNL-RHIC, and CERN-LHC.\\[0.4cm]
$^*$Dedicated to our mentor Walter Greiner; to be published in the memorial volume edited by Peter O. Hess.

\end{abstract}


\body

\section{Introduction}

Just fifteen years after the coincident creation in 1964 of two great ideas governing the strong interactions ---  quarks and the Hagedorn temperature \Th --- these two concepts merged, giving birth to a new discipline, the physics of the novel fifth state of matter, the quark-gluon plasma (QGP). Today there is consensus that QGP filled the cosmos during the first 20\,$\mu$s after the Big-Bang. For  three decades laboratory experiments at the European Center for Particle Physics (CERN) and Brookhaven National Laboratory (BNL) have been exploring this primordial phase of matter colliding nuclei at relativistic energies.

As the ideas about QGP formation in \RHI\ collisions matured a practical challenge  emerged: How can the locally color deconfined QGP  state  be distinguished from a gas of confined hadrons? In the period 1979--86 the strangeness signature of QGP was developed for this purpose, with our 1986 review~\cite{Koch:1986ud} in essence completing the theoretical foundations. In a review of 1995 one of us (BM) presented a compendium of possible QGP signatures.~\cite{Muller:1994rb}.

During this period and in the following years many scientists were wondering whether QGP could be detected as a matter of principle. Could it be that a quark-gluon based description is merely a change of the Hilbert space basis, \ie\ a unitary change between quark and hadron bases? If so, maybe there would be nothing to be discovered! 

A globally color-singletfireball composed of quarks and gluons and several Fermi in diameter is, in principle, a hadron, \ie, a strongly interacting object. Today it is also understood that for an infinite QCD system there is no discontinuity in the equation of state of the baryon symmetric QCD matter. But the key to understanding why the QGP is a physically meaningful and observable concept is to ask the following question: Do hadronic states exist in nature containing many more than three quarks which cannot be factorized into color-singlet components, each containing a few quarks? Even if the complete Fock space of hadrons includes an extended QGP, such a state is distinct from states, such as ordinary atomic nuclei, in which color-singlet hadrons containing few quarks propagate across arbitrary distances while colored quarks and gluons are confined inside such locally color-singlet hadrons. 

BM presented these thoughts at the {\em Quark Matter 1991} conference in Gatlinburg.~\cite{Muller:1991jk} The ensuing heated discussions reverberate in retrospect, as they lie at the core of the lingering doubts about the observability of QGP formation in \RHI\ collisions that remained widespread even among experts for many years. As we discuss below, this misunderstanding was, in part, exacerbated by the fact that many observables that were proposed as QGP \lq\lq signatures\rq\rq\ are not sensitive to this defining property of the QGP, \ie, local color deconfinement.

The discussions at {\em Quark Matter 1991} are recalled here as a reminder against which odds the experimental efforts aimed at discovering the QGP had to struggle and to explain why the discovery of QGP, which had in fact occurred at the time of that conference, (a) was not recognized as valid by one of the discovering experimental groups; (b) needed to wait nine more years to be announced by the CERN Laboratory where the experiments were being performed; (c) had to wait fourteen additional years before a competing laboratory, BNL, concurred after an intense intellectual struggle; and (d) is still a sometimes disputed discovery a quarter century later. 

We describe below the pivotal CERN-SPS experiments that, to the apparent disbelief of some of the involved scientists (see Subsect.~\ref{ref:NA35}), created the QGP phase of matter (Subsect.~\ref{ref:WA85}). In the following years these results found their confirmation in the Pb+Pb collision program at the CERN-SPS (Subsect.~\ref{ssec:CERNQGP}), leading to the CERN February 2000 QGP announcement based on two campaigns of experiments and many refereed and published articles. The QGP discovery was confirmed five years later by experiments at the BNL Relativistic Heavy Ion Collider (RHIC) employing new, independent probes that helped establish a broad public consensus (Sect.~\ref{sec:BNL}), and later by further experiments at CERN, both at SPS and LHC (Subsect.~\ref{ssec:QGPafter}). 

The strangeness signature, which enabled the first clear observation of the QGP was originally conceived and proposed by one of us (JR) at CERN and later developed to full maturity by us initially in Walter Greiner's Institute. Walter was at that time among the vocal skeptics of our work and of QGP research more generally. However, his broad principled opposition to the subject provided additional inspiration for our work, which continued at the University of Cape Town. 

In hindsight, it is puzzling why Walter originally considered all QGP research with such skepticism, because he generally loved innovative, even \lq\lq exotic\rq\rq\ physics. He became, for example, a strong advocate of searches for stable or meta-stable multi-strange cold quark drops called strangelets. Whatever the reasons may have been, in later years Walter's institute, with his strong support, became the preeminent German center for theoretical QGP physics, and today QGP is a core component of the research program of the Frankfurt Institute for Advanced Study (FIAS), which he co-founded.

\section{Strangeness: The pivotal QGP signature}

The existence and observability of a new phase of elementary matter, the QGP, must be demonstrated by experiment. This requires identification of probes of QGP that are:
\begin{enumerate}
\item operational on the collision time scale of $10^{-23}$\,s;
\item sensitive to the local color charge deconfinement allowing color charges to diffuse freely throughout the matter;
\item dependent on the gluon degree of freedom, which is the characteristic new dynamical degree of freedom.
\end{enumerate}

The heaviest of the three light quark flavors, strangeness, emerged 1980--82 as the pivotal signature of QGP satisfying these three conditions. When color bonds are broken, the chemically equilibrated deconfined state contains an unusually high abundance of strange quark pairs~\cite{Rafelski:1980rk,Rafelski:1980fy}. This statistical argument was soon complemented by a study of the dynamics of the strangeness (chemical) equilibration process. We found that predominantly the  gluon component in the QGP  produces strange quark pairs rapidly, and just on the required time scale.~\cite{Rafelski:1982pu} Our work also connected strangeness enhancement to the presence of gluons in the QGP. The high density of strangeness at the time of QGP hadronization was a natural source of multi-strange hadrons~\cite{Rafelski:1982ii}, if hadronization proceeded predominantly by the coalescence of pre-existing quarks and antiquarks~\cite{Koch:1986ud}.

By Spring 1986 we had developed a detailed model and presented predictions showing how the high density and the mobility of already produced strange and antistrange quarks in the fireball favors the formation of multi-strange hadrons during hadronization~\cite{Koch:1986ud}. We also showed that these particles are  produced quite rarely if only individual hadrons collide. We presented  a detailed discussion of how a fireball of deconfined quarks turns into strangeness carrying hadrons and showed that multi-strange antibaryons are the most characteristic signatures of the QGP nature of the source. By distinguishing the relative chemical equilibrium from the absolute yields of quark pairs we introduced what today is called the statistical hadronization model  which allows us to measure the chemical properties of the hadron source~\cite{SHAREa,SHAREb,SHAREc}. 

Though the production of final state hadrons characterizes the conditions in the QGP fireball at the time of its breakup (hadronization) the total strangeness flavor  yield provides in situations when chemical equilibrium in QGP fireball is hard to achieve additional  information about conditions arising in the first instants of the reaction. In this sense strangeness alone, when studied in depth, can provide a wealth of insights about the formation and evolution of the QGP fireball. As we discuss below there are other observables available to explore the properties of the early stage QGP.  

Considering all produced hadronic particles it is possible to evaluate the property of the  dense matter fireball. A fireball of QGP that expands and breaks apart should do this  in a manner that does not remember in great detail the mechanisms that led to the formation of the thermal fireball. Indeed, one of the important findings emerging from studies of the hadronization process is that the hadron chemical freeze-out conditions are universal~\cite{Letessier:2005qe,Petran:2013qla,Rafelski:2015cxa}. This universality is further consistent with the  sudden hadronization mechanism we first studied 30 years ago~\cite{Koch:1986ud}.

The strangeness observable was and is experimentally popular since strange hadrons are produced abundantly and can be measured over a large kinematic domain. Therefore, a large body of experimental results is available today. All of these results are consistent with hadronic particle production occurring from a dense source in which the deconfined strange quarks are already created before hadrons are formed. These (anti-)strange quarks are free to move around or diffuse through the QGP and are readily available to form hadrons. 

Once one has confirmed that a QGP was formed, other observables can be interpreted on that basis. However, few, if any, other QGP observables probe the characteristic nature of the source in a way that would uniquely pinpoint a QGP  with local color deconfinement at the time of hadron formation. Let us discuss a few examples:
\begin{itemize}
\item
\underline{Fluid dynamics}: The fireball is recognized to consist of matter described by hydrodynamical simulations,~\cite{Bjorken:1982qr,Ollitrault:1992bk} which implies the fireball is comprised of a near minimal specific viscosity liquid.~\cite{Song:2010mg}. It is natural to associate this result with a fluid composed of relativistic, strongly interacting particles \eg\ quarks and gluons, but does not by itself signal that the fluid is a QGP. Indeed, it is still not entirely clear how the QGP acquires its nearly \lq\lq perfect\rq\rq\ liquid nature at thermal length scales.
\item
\underline{Jet quenching} is observed in a clear and convincing way in \RHI\ collisions at sufficiently high energy.~\cite{Bjorken:1982tu,Wang:1991xy,Adcox:2001jp,Chatrchyan:2011sx} Arguably, this property signals the formation of a fireball endowed with a high density of color fields, which impedes the escape of high energy particles (Subsect.~\ref{ssec:jets}).
\item
\underline{Quarkonium production} can occur in primary collisions~\cite{Shuryak:1978ij} and charm  recombinant hadronization~\cite{Thews:2000rj}. Heavy quarkonia, like jet emission, can be suppressed~\cite{Matsui:1986dk}  in interaction with dense matter. Therefore the yield is determined by the interplay of at least three different mechanisms with processes contributing needing to be modeled in detail. Today, the increased charmonium yield at the LHC caused by $c\bar{c}$ recombination, similar to the enhanced formation of multi-strange baryons observed over a wide energy range, is often considered as the most convincing quarkonium signature of QGP formation (Subsect.~\ref{ssec:QQ}).
\item
\underline{Electromagentic signals}: Photons and dileptons are the most penetrating probes promising insights into the initial dynamics of QGP formation and evolution.~\cite{Feinberg:1976ua,Shuryak:1978ij,Kapusta:1991qp} These observables will without doubt come of age in the future high luminosity RHIC and LHC runs. Today we can use them in a semi-quantitative manner to estimate, \eg, the initial temperature of the fireball (Subsect.~\ref{ssec:evolution}).
\end{itemize}

In the next Section we briefly recapitulate the properties and evolution of a thermal QGP fireball formed in \RHI\ collisions, followed by a review of the chronology of the strangeness signature providing evidence for QGP formation in Section \ref{sec:str}. There we describe the initial CERN-SPS research program, the first series of experiments, and the more mature Pb+Pb collision experiments that reconfirmed the early observations and led to the February 2000 CERN announcement of the \lq\lq discovery\rq\rq\ of the QGP, as well as more recent strangeness developments. The QGP discovery announcement by the RHIC community relied on other observables and is described in Section~\ref{sec:BNL}.


\section{Quark-Gluon Plasma}\label{sec:QGP}
\subsection{Evolution of fireball in time}\label{ssec:evolution}

In  laboratory experiments involving collisions of large nuclei at relativistic energies, several (nearly)  independent reaction steps occur and ultimately lead to hadron production:
\begin{enumerate}
\item Formation of  the primary fireball; a momentum equipartitioned partonic phase comprising in a compressed space-time domain most of the final state entropy;
\item The cooking of the energy content of the hot matter fireball towards chemical (flavor) equilibrium in a hot QGP phase;
\item Emergence of transient massive quarks due to spontaneous chiral symmetry breaking and disappearance of free gluons; from this point on the fireball cannot in chemical equilibrium if entropy, energy, baryon number, and strangeness are to be conserved;
\item Hadronization, \ie\ is the coalescence of effective and strongly interacting up, down, and strange quarks and anti-quarks   into the final state hadrons, with the coalescence probability weighted by accessible phase space. The hadronization process can be subject to detailed experimental study, resulting in determination of the physical properties and statistical parameters that govern the process. We discuss this below.
\end{enumerate}

Experimental information about the maximum temperatures reached in heavy ion collisions can be derived from the spectrum of radiated photons in the energy range $E_T = 1-3$ GeV, where direct photon emission is dominated by thermal radiation. The analysis of the measured spectrum in terms of the thermal properties of the fireball is somewhat model dependent, since the observed yield spans the entire collision history. In practice, the hydrodynamical simulations of the time evolution set a lower limit to the initial temperature. In Au+Au collisions at RHIC ($\sqrt{s_{\rm NN}} = 200$ GeV), this initial temperature exceeds 300~MeV; in Pb+Pb collisions at the LHC ($\sqrt{s_{\rm NN}} = 2.76$ TeV) the initial temperature is at least 450~MeV. This increase is in good agreement with the observed scaling of the particle multiplicity from RHIC to LHC, which indicates a substantial  increase   in the fireball entropy content. 

\subsection{QCD Matter in heavy ion collisions}

Lattice gauge theory has made impressive progress on the calculation of static thermodynamic properties of baryon symmetric QCD matter. The equation of state  for physical quark masses is now known for  $\mu_\mathrm{B} \lesssim \mu_\mathrm{B,cr}$  with precision. The quasi-critical temperature where susceptibilities related to chiral symmetry peak has been determined in lattice QCD simulations:~\cite{Borsanyi:2012rr} $T_c(\mu_\mathrm{B}=0) \approx 150$ MeV. A critical point is expected in the $\mu_\mathrm{B}$ domain  that can be explored in experiments carried out at SPS and RHIC-BES, as is illustrated in left-hand part of \rf{QGPhase}, which shows the approximate range of $T$ explored in RHIC and LHC experiments with the corresponding energy density versus temperature at $\mu_\mathrm{B}=0$ curve calculated by lattice QCD. 

\begin{figure}
\centerline{
\includegraphics[width=0.46\linewidth]{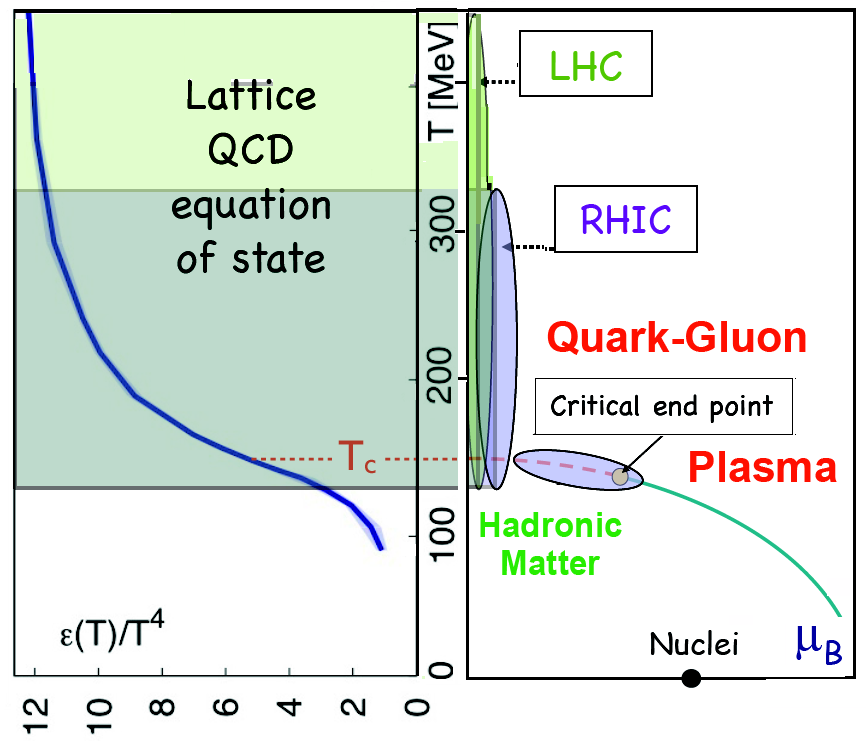}
\includegraphics[width=0.54\linewidth]{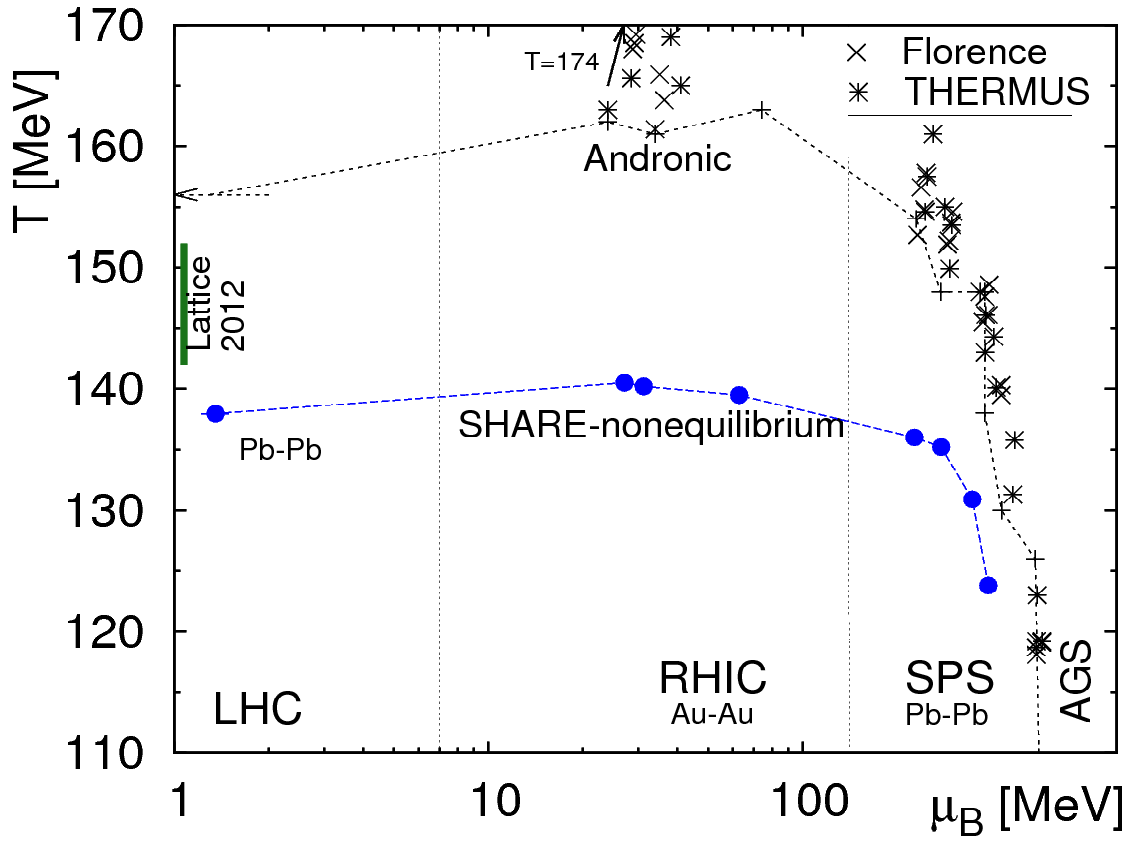}
}
\caption{On left: Phase diagram of QCD matter (right part of left panel) overlaid with regions covered by LHC and RHIC. The experimentally covered ranges are projected onto the energy density versus temperature at $\mu_\mathrm{B}=0$ curve calculated by lattice QCD (left part of left panel). The current status of statistical hadronization model analysis showing different results obtained for the chemical freeze-out points in the $T$-$\mu_\mathrm{B}$ plane, update of results shown in Ref.~\cite{Rafelski:2015cxa}\label{QGPhase}}
\end{figure}
On right in \rf{QGPhase} we show how these results stack up against the results obtained across   20 years of hadron chemical freeze-out points; chemical freeze-out is where the yields of produced hadrons will not change and hence these results \underline{must be} below quasi-critical temperature. We see two lines describing results of statistical hadronization model final state analysis that span a range of $\mu_\mathrm{B}$ fitted to the data in SPS, RHIC and LHC experiments. In between these two is the Lattice QCD critical temperature obtained and reconfirmed in past 5 years. From these results emerges clearly that the chemical nonequilibrium description of hadron production by a QGP is the only model compatible with the current understanding of strongly interacting matter.  Note that the curve marked SHARE  assumes as in our  1986  work relative chemical equilibrium, but allows the freedom in the yields of all quark-pairs. For further details see   Ref.\,\cite{Rafelski:2015cxa}.

\subsection{Properties of QCD matter}
It is worthwhile asking which intrinsic properties of the quark-gluon plasma medium dynamics we can hope to determine experimentally and from which observables. A not exhaustive list seen already in the 1992 presentation~\cite{Muller:1991jk} includes:
\begin{itemize}
\setlength{\itemsep}{0pt}
\item 
The equation of state of the matter, given by relations among the components of the energy-momentum tensor $T_{\mu\nu}$ at equilibrium and their temperature dependence are reflected in the spectra of emitted particles. Lattice QCD is able to compute these quantities reliably. The analysis of chemical freeze-out conditions provides pressure $P$, energy density $\epsilon$ and entropy density $\sigma$.\cite{Rafelski:2015cxa}
\item
Transport coefficients of the quark-gluon plasma, especially the shear viscosity $\eta$, the coefficient $\hat{q}$ governing the transverse momentum diffusion of a fast parton (often called the {\em jet quenching parameter}), the coefficient of linear energy loss $\hat{e}$, and the diffusion coefficient $\kappa$ of a heavy quark, are related to the final-state flow pattern and the energy loss of fast partons that initiate jets. Lattice gauge theory presently cannot reliably calculate these dynamical quantities.
\item 
The static color screening length $\lambda_D$ (the inverse Debye mass $m_D$) governs the dissolution of bound states of heavy quarks in the quark-gluon plasma. This static quantity can be reliably calculated on the lattice.
\item
The electromagnetic response function of the quark-gluon plasma is reflected in the emission of thermal photons and lepton pairs. This dynamical quantity is difficult to calculate on the lattice, but moderate progress has been made recently.
\end{itemize}
All but the last of these properties are microscopically related to correlation functions of the gauge field. This implies that the associated experimental observables are mostly sensitive to the gluon structure of the quark-gluon plasma. On the other hand, much more is known theoretically from lattice simulations about the quark structure of hot QCD matter, because it is much easier to construct operators from quark fields that can be reliably calculated. In this respect, lattice calculations and heavy ion experiments are to a certain degree complementary. The presence of jets in heavy ion collisions at LHC and RHIC tell us that at high virtuality $Q^2$ or high momenta $p$, the QGP is weakly coupled and has quasi particle structure. On the other hand, the collective flow properties of the matter produced in the collisions tells us that at thermal momentum scales the quark-gluon plasma is strongly coupled. At which $Q^2$ or $p$ does the does the transition between strong and weak coupling occur? Does the quark-gluon plasma still contain quasi particles at the thermal scale?  Which observables (jets?) can help us pinpoint where the transition occurs?

Leaving aside these theoretical contemplations we now turn to the empirical observable, the strangeness content of QGP fireball. The reason that this is of interest is the undisputed observation that the yield  of strange quark pairs is noticeably greater in \RHI\ collisions compared to typical high energy elementary particle reactions. How this happens was addressed in our 1986 review~\cite{Koch:1986ud}. In the following we focus on  what happens next: how this medium heavy flavor hadronizes, what type of particles are produced, and what this tells about the physics of the source, so that our discussion can concentrate on the physics of the fireball source as observed through the \lq\lq eye\rq\rq\ of strangeness.

\section{Probing QCD Matter with Strangeness}\label{sec:str}
\subsection{CERN-SPS experiments: Overview}\label{ssec:SPS}

The \rf{CERNhadrons} shows the time line of those CERN experiments conceived in the early-to-mid 1980s that contributed to the observation of strangeness by means of study of emitted hadrons. Note that CERN accelerated oxygen ($^{16}$O) ion beams at 200 \agev  in 1986, followed within a year by sulfur ($^{32}$S) beams at the same energy, and in the mid-1990s by lead ($^{208}$Pb) beams at 158 \agev. First results from Pb+Pb collisions were reported in 1996; this is indicated on the left in  \rf{CERNhadrons}. We will address below the pivotal results obtained by the two experimental series WA85--NA57 and NA35--NA49. 
\begin{figure}
\centerline{%
\includegraphics[width=0.95\columnwidth]{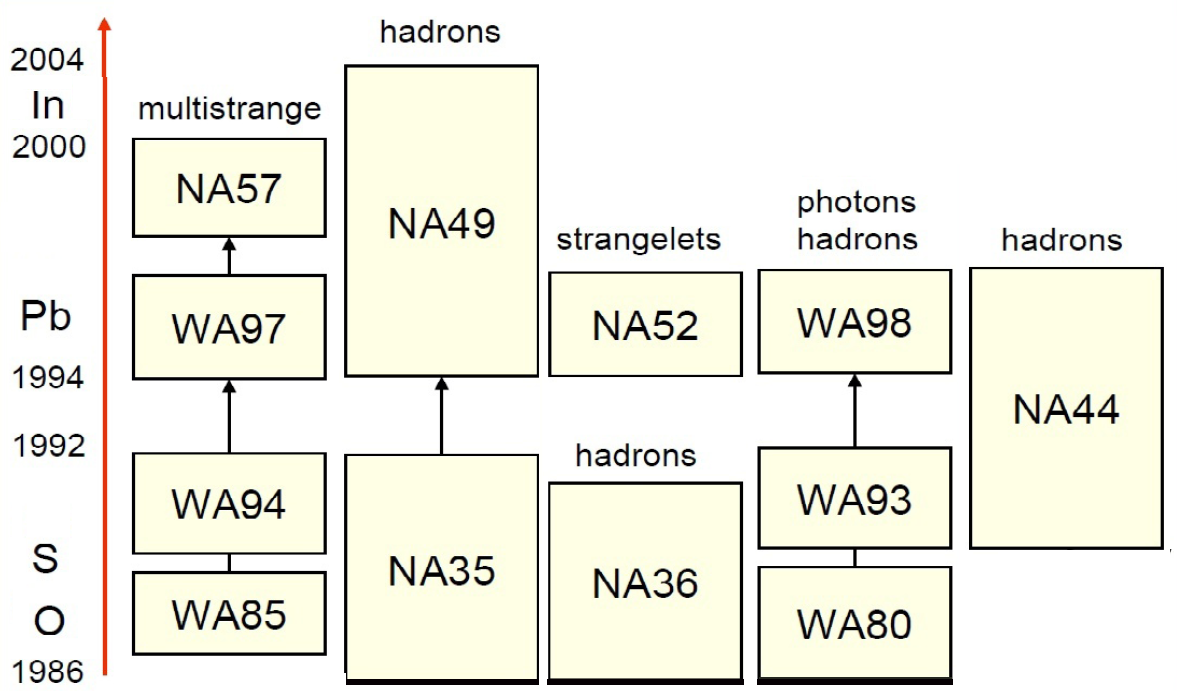}
}
\caption{The multitude of CERN experiments measuring hadron production  on line prior to the year 2000 QGP announcement.}
\label{CERNhadrons}
\end{figure}

The WA85 experiment started a bit later than the other Day-1 experiments, and it is instructive to understand the reasons for this delay: The initial experimental proposal was directed at our strange antibaryon signature of QGP; however, the CERN-SPS advisory committee was influenced by arguments by senior theorists that the strange antibaryon enhancement  could not be observed since, when QGP converts into hadrons, the multi-strange antibaryon ($\overline{\Xi}$ and $\Omega$, $\overline {\Omega}$) signature would be erased by annihilation in the baryon-rich environment produced in collisions at the SPS energy. 

As  we know today, the assumptions we made in our work are, in fact, realized in nature. The QGP fireball explodes so rapidly after hadronization occurs that many hadrons are produced free-streaming into the vacuum, and chemical reactions involving strangeness quickly fall out of equilibrium during the explosive expansion. Strange antibaryon annihilation does not prevail. This can easily be seen by comparing baryon and antibaryon $p_\bot$-spectra as we show below. Another evidence for the sudden hadronization process, as this became known, is that the yields of hadrons follow the predicted pattern imposed by entropy, baryon number, strangeness content of QGP fireball, which means that the hadron yields follow a scenario today called \lq\lq chemical non-equilibrium\rq\rq.

Despite this criticism the experiment WA85 was eventually approved in Fall 1986, (a) because a way was found to modify the CERN $\Omega$-spectrometer to observe less exotic strange particles including kaons $(\bar s q),\, (\bar q s)$ and singly strange antihyperons $\overline{\Lambda}(\bar q\bar q\bar s)$; and (b) on the strength of the arguments presented in our 1986 published work~\cite{Koch:1986ud}.

\subsection{CERN-NA35 Experiment}\label{ref:NA35}

The CERN-NA35 experiment was an extension of the LBNL-GSI collaboration at the BEVALAC with LBNL's Howell Pugh being the main force for strangeness. Howell was a member of both the NA35 and the NA36 experiments; however, NA36 was highly advanced and had instrumental difficulties, whereas NA35 relied on well established technology. Initially the objective of NA35 was the exploration of the equation of state of dense nuclear matter, a direct continuation of the effort carried out at BEVALAC by most of those involved in the experiment. The fade-out of the strangeness focused NA36 presented the NA35 experimental program with the opportunity to expand into the strangeness signature of QGP.  
 
Writing in 2000, Gra\.zyna Odyniec~\cite{Odyniec:2001} of LBNL commented \lq\lq From the very beginning Howell [Pugh], with firmness and clarity, advocated the study of strange baryon and antibaryon production. He played a leading role in launching two of the major CERN heavy-ion experiments: NA35 and NA36, the latter being exclusively dedicated to measurements of hyperons. Strangeness enhancement predicted by theorists was discovered by NA35 and reported at {\em Quark Matter 1988}.\rq\rq\  The NA35 results were presented in their extended and final form in 1990. The published article~\cite{Bartke:1990cn} stated in its abstract: \lq\lq Significant enhancement of the multiplicities of all observed strange particles relative to negative hadrons was observed in central S+S collisions, as compared to $p+p$ and $p+$S collisions.\rq\rq\ In the concluding section the authors commented: \lq\lq Thus our observation \ldots appears to be consistent with a dynamical evolution that passes through a deconfinement stage.\rq\rq\ 

Yet the article refrained from making a discovery claim by continuing: \lq\lq However, ... this may not be the only explanation because the possible pre-equilibrium aspects of the early interpenetration stage, or even the conceivable overall off-equilibrium nature of the entire dynamics, may also lead to enhanced strangeness production, even without plasma formation.\rq\rq\ In retrospect, this statement appears puzzling, because it had already been shown five years earlier~\cite{Koch:1984tz}  that the strangeness production and evolution within the hadronic gas phase could not chemically equilibrate strangeness. In their {\em Quark Matter 1990} proceedings~\cite{Baechler:1991pp} report the NA35 Collaboration distanced itself even further from a possible QGP discovery claim by stating: \lq\lq We have demonstrated a two-fold increase in the relative $s + \bar s$ concentration in central S-S collisions, both as reflected in the $K/\pi$ ratio and in the hyperon multiplicities. A final explanation in terms of reaction dynamics has not been given as of yet.\rq\rq\  

Clearly, the NA35 collaboration missed a unique opportunity to claim first observation of our QGP strangeness signature by refraining to interpret their experimental results in these terms. In the following, the publications of NA35 strangeness results regularly side-stepped any explicit mention of QGP formation as an explanation of their observations. It is tempting to speculate why this occurred: A motivation may have been that the idea of QGP formation in collisions of mid-sized nuclei in the SPS energy range ran counter to the views of many members of the high energy physics community, some being members of NA35, who thought that only RHIC with its ten-fold higher collision energy would be capable of creating a QGP.  We leave further discussion of this question to future historians of science. 

\subsection{CERN-WA85 Experiment}\label{ref:WA85}

Against this background the late arriving CERN  $\Omega\rq$ spectrometer experiment WA85 under the leadership of Emanuele Quercigh took center stage of  QGP search at SPS with published results on $\Lambda$ and $\bar{\Lambda}$~\cite{Abatzis:1990cm}, $\Xi^-$, $\overline{\Xi^-}$~\cite{Abatzis:1990gz}, and a systematic exploration of the parametric dependence of both observables, showing characteristics of a QGP~\cite{Abatzis:1991ju}. In contrast to NA35, the WA85 Collaboration takes a much firmer position, citing evidence in favor of QGP formation: \lq\lq The(se) results indicate that our $\overline{\Xi^-}$ production rate, relative to $\bar{\Lambda}$, is enhanced with respect to $pp$ interactions; this result is difficult to explain in terms of non-QGP models [11] or QGP models with complete hadronization dynamics [12]. We note, however, that sudden hadronization from QGP near equilibrium could reproduce this enhancement~[2].\rq\rq\ 

Ref.\,[2] is an analysis~\cite{Rafelski:1991rh} of these WA85 results within the nascent Statistical Hadronization Model  published in March 1991 by one of us (JR). In this work strange baryon and antibaryon particle production data for S--W collisions were used to determine the \lq chemical\rq\ properties of the particle source, \ie the chemical potentials and phase space occupancy. In the abstract of this analysis we read: \lq\lq Experimental results on strange anti-baryon production in nuclear S--W collisions at $200 A$\,GeV are described in terms of a simple model of an explosively disintegrating quark-gluon plasma (QGP).\rq\rq\ The summary closes with, \lq\lq We have presented here a method and provided a wealth of detailed predictions, which may be employed to study the evidence for the QGP origin of high $p_\bot$ strange baryons and anti-baryons.\rq\rq\ 

\begin{figure}
\centerline{%
\includegraphics[width=6.5cm]{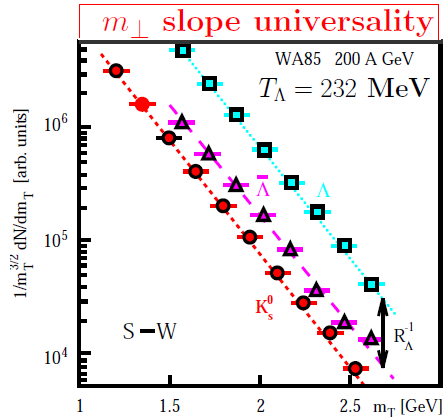}\hspace*{0.5cm}
\includegraphics[width=4.6cm]{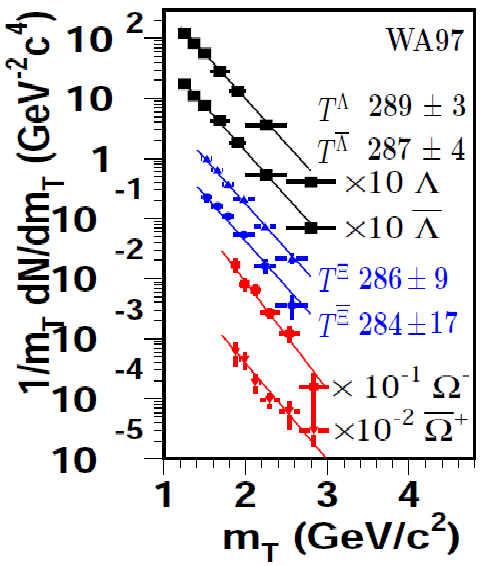}
}
\caption{The universality of baryon-antibaryon $m_\bot$ spectra obtained by WA85 (on left), and by WA97 (on right) demonstrate that baryons and antibaryons  are produced in the same explosive manner by the fireball of dense QGP.}
\label{AntiHymT}
\end{figure}

The WA85 paper~\cite{Abatzis:1991ju} cited above echoed this point of view, leading the WA85 collaboration in early 1991 to claim the QGP discovery. From today's vantage point, we can say that with this 1990/91 analysis method and the WA85 results and claims of the period, the QGP had been discovered; however, the discovery was not universally accepted by the broader physics community, not even by the majority of the community involved in \RHI\ experiments at the CERN-SPS. For a more extensive discussion of the WA85 results and the arguments behind the discovery claim we refer to the popular review by the spokesman of the WA85 Collaboration, Emanuele Quercigh, prepared with JR.\cite{QR2000}.

\subsection{The path to CERN QGP discovery announcement}\label{ssec:CERNQGP}

Strangeness results continued to be published by these two CERN experimental groups: NA35/NA49 (evolving into NA61)  and WA85/WA94/WA97 (evolving into NA57) in the ensuing decade through the early 21$^{\rm st}$ century. The new experiment NA49 that replaced NA35 was evolving now. When CERN announced in early February 2000 the discovery of a new phase of matter, this event followed a decade of experimental work with dozens of refereed papers published showing agreement of QGP strangeness signature with our QGP based predictions.   

\begin{figure}
\centerline{%
\parbox{6.2cm}{\includegraphics[width=6cm]{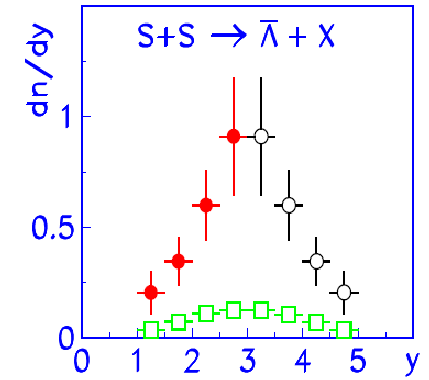}}
\parbox{6.cm}{\includegraphics[width=5.8cm]{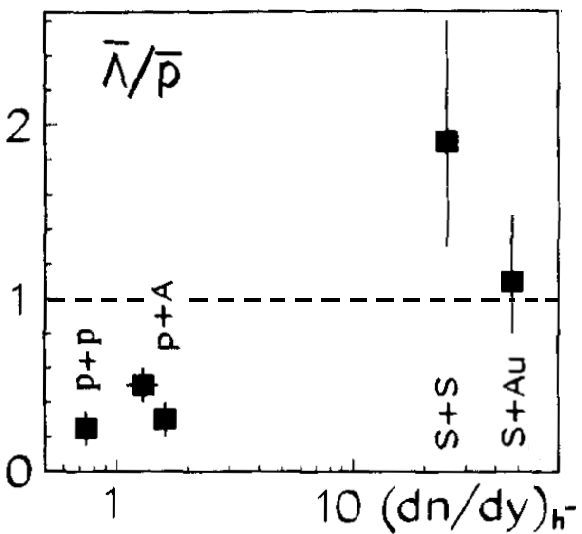}}}
\caption{NA35-antihyperon results for 200 \agev S-A collisions of July 1995. Left: Distribution in rapidity of $\bar\Lambda$ for S-S compared to the yield in $pp$ collisions scaled up with relative abundance of negatives $h^-$, from Y. Foka thesis~\cite{Foka:1995Thesis}; Right: $\bar\Lambda/\bar p$ ratio as a function of mean $h^-$ multiplicity; results adapted from Ref.\,\cite{Alber:1996mq}..\label{AntiHypFig}}
\end{figure}

The NA35 Collaboration presented the ratio $ \overline{\Lambda}/ \bar p\sim 1.4$ measured near mid-rapidity in Summer 1995~\cite{Alber:1996mq}, showing a three-to-fivefold enhancement, dependent on the collision system as compared to measurement in more elementary reactions. This was the QGP signature of the first strangeness papers in 1980~\cite{Rafelski:1980rk,Rafelski:1980fy}. Due to the shift of the central rapidity for asymmetric collisions the decrease in this ratio as the asymmetry increases is in agreement with theoretical expectations; these results are shown on the right in \rf{AntiHypFig}.  

The WA85/94 collaboration focused on multi-strange baryon and antibaryon ratios, for $ \overline{\Xi}/\overline{\Lambda}$ see \eg\ the 1993 review of David Evans~\cite{Evans:1994sg}. A full summary of all results is contained in the review of Federico Antinori of 1997~\cite{Antinori:1997nn} and shown in \rf{WA85WA94AntiHFig} with data referring to the WA85/94 reports presented at the {\em Quark Matter 1995} conference~\cite{DiBari:1995cy,Kinson:1995cz}. 

\begin{figure}
\centerline{%
\includegraphics[width=0.8\linewidth]{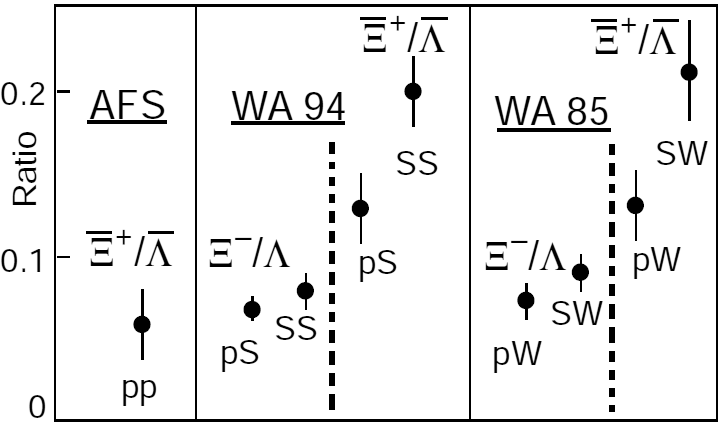}}
\caption{WA85 and WA94 hyperon and antihyperon results for 200 GeV/A S-A collisions of January 1995, adopted from~\cite{Antinori:1997nn}. \label{WA85WA94AntiHFig}}
\end{figure}

Seeing these developments one could not help but being convinced that the then forthcoming Pb+Pb experiments at CERN would confirm and cement the strangeness based QGP discovery before the end of the old millennium. This was indeed the case as we see, for example, in the retrospective summary \rf{AliceALL}: in this 2013  strangeness enhancement review by the CERN-ALICE collaboration we see that the Pb+Pb CERN-SPS antihyperon enhancement at $\sqrt{s_\mathrm{NN}}=17.2$ GeV (equivalent to\,158\,\agev) reaches the value 20 for the triply strange $\Omega+\overline\Omega$ baryons (open triangles). This is the most dramatic medium modification result ever recorded in \RHI\ collisions. 

\begin{figure}[!b]
\centerline{%
\includegraphics[width=0.8\linewidth]{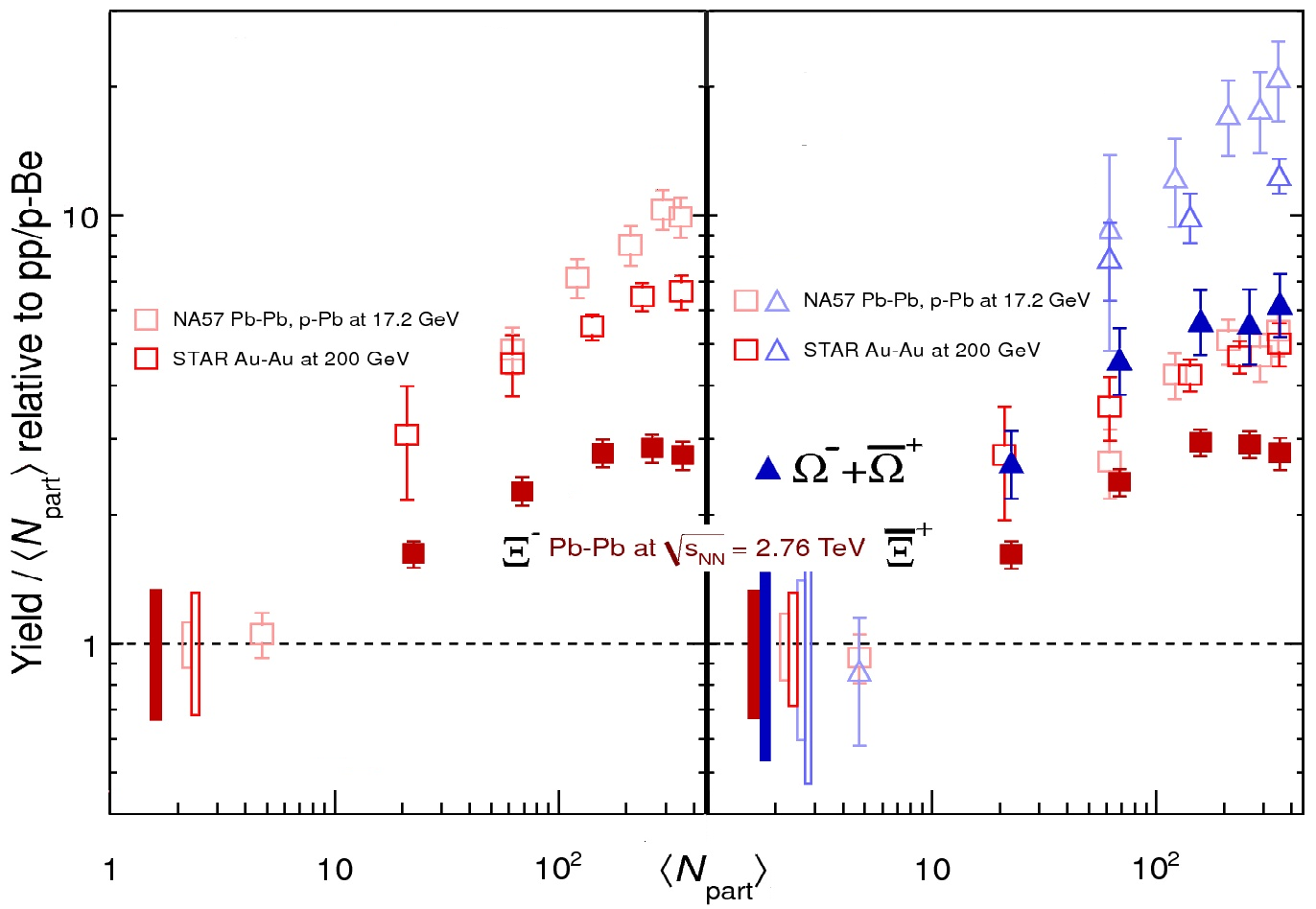}}
\caption{ALICE-LHC, STAR-RHIC, and NA57-SPS $N_\mathrm{part}$ normalized hyperon (left), and antihyperon (right) yields as function of $N_\mathrm{part}$; yields renormalized arbitrarily to unity for smallest $N_\mathrm{part}$ available, adopted from~\cite{ABELEV:2013zaa}.\label{AliceALL}}
\end{figure}

Despite these impressive (anti-)hyperon results available at the time of the CERN QGP announcement~\cite{Heinz:2000bk} in February 2000, the announcement was based on a  \lq\lq consensus\rq\rq\ presentation of all heavy-ion SPS-CERN physics results. There was a strategic problem, since some  of these experiments had results that had little to do with the QGP discovery itself or were statistically marginal.  With the \lq\lq consensus\rq\rq\ strategy CERN opened its QGP discovery up to criticism, where both the interpretation of the data and the statistical validity of some of the results could be questioned. This problem spilled over to the strangeness signature which should have been uncontested on both grounds. If CERN management was not willing to base its claim of QGP discovery on the impressive observations of strangeness enhancement alone, there must exist a scientific reason for this lack of confidence! 

Would an announcement of the QGP discovery been more compelling, if it had been based solely on the strangeness signature? While it is tempting to think so, it probably was not practical, because a very sizable fraction of the \RHI\ physics community questioned then, and has continued to question, the relevance of strangeness as a signature of the QGP, in spite of the fact that strangeness enhancement is the sole QGP signature for which every single prediction has been quantitatively confirmed experimentally and no comprehensive alternative explanation has been given.

In summary, we firmly believe that CERN had a justifiable case for the discovery of QGP with SPS results addressing strangeness and multi-strange antihyperon production. However, the lack of consensus within the community precluded this simple approach and the global \lq all experiments\rq\ claim advanced by CERN was not convincing to outside observers.  It took another experimental program, that of the Relativistic Heavy Ion Collider commencing in 2000, with a broader range of accessible observables to create a wider consensus, leading up to a second announcement of QGP discovery in 2005.

\subsection{Developments at CERN after 2000}\label{ssec:QGPafter}

The NA49 collaboration evolving into NA61 focused its main objective since the CERN QGP discovery announcement on the search of a threshold in beam energy for the previously observed phenomena. In a systematic experimental study of the K$^+/\pi^+$ ratio~\cite{Gazdzicki:2010iv} as function of energy the so-called \lq\lq Marek's Horn\rq\rq\ was discovered, named after Marek Gaz\rq dzicki, now spokesman of the NA61 collaboration.  

The Statistical Hadronization Model analysis of this feature is seen in \rf{Kpi2008Fig} on the left, where the K$^+/\pi^+$-ratio for both experiment and theoretical fit adapted from Ref.\,\cite{Rafelski:2009gu} is shown. The question that we need to answer is what it is that the  \lq\lq horn\rq\rq\ signals? To this end we look at the right-hand panel of \rf{Kpi2008Fig} showing the ratio of the thermal fireball energy to the number of strange quark pairs $E/s$. We clearly see that at the location of the horn-like feature the energy cost of producing a pair of strange quarks is leveling off at a low value, signaling onset of a new, more energy efficient, production mechanism. 

\begin{figure}[bht]
\centerline{%
\includegraphics[width=4.8cm]{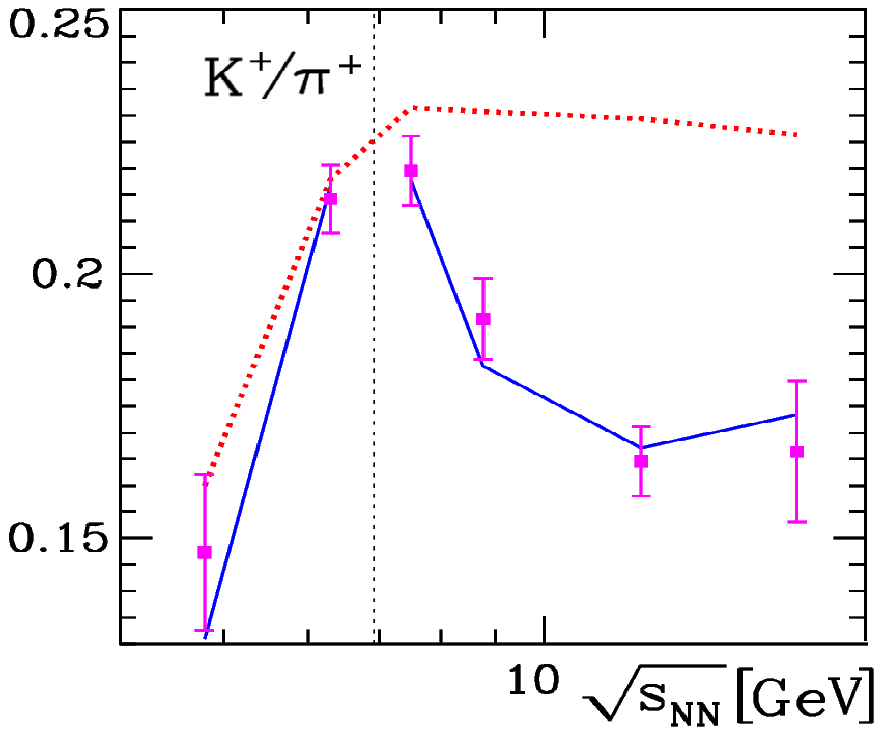} 
\includegraphics[width=7.0cm]{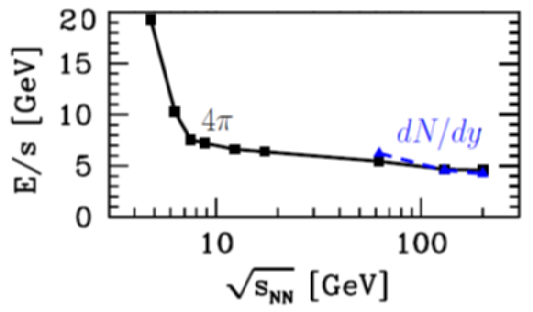}}
\caption{\underline{Left:} K$^+/\pi^+$-ratio adapted from Ref.\,\cite{Rafelski:2009gu}; AGS (lowest $\sqrt{s_\mathrm{NN}}$) and NA49-SPS energy range results are connected by the (blue)   line indicating at the edges the theoretical fit result within chemical nonequilibrium model, the dotted line shows best chemical equilibrium result.  \underline{Right:} Cost in fireball thermal energy of a strangeness pair, $E/s$ as a function of CM collision energy $\sqrt{s_\mathrm{NN}}$. $4\pi$ results (black) are estimates for RHIC, line guides the eye; RHIC domain (blue) shows $(dE/dy)/(ds/dy)$. Update of result of Ref.\,\cite{Letessier:2005qe}.\label{Kpi2008Fig}}
\end{figure}

This $E/s$ curve in the right panel of \rf{Kpi2008Fig} is derived from the fit to the observed particle yields which fit the \lq\lq horn\rq\rq\  but needs further interpretation. Considering our prior extensive work showing that quark based processes are less effective than gluon mediated processes in producing strange quark pairs, we conclude that in the knee of  $E/s$ glue degrees of freedom must have been fully activated. The conclusion is that QGP has been formed at collision energies above $\sqrt{s_\mathrm{NN}}=7$\,GeV. Below this threshold we see a transition domain where with decreasing energy more and more fireball energy is required to make strange quark pairs.  Future experiments (the second RHIC Beam Energy scan will probe down to $\sqrt{s_\mathrm{NN}}=3$\,GeV) will tell if this rise saturates at lower collision energies.

\begin{figure}
\centerline{%
\includegraphics[width=0.7\linewidth]{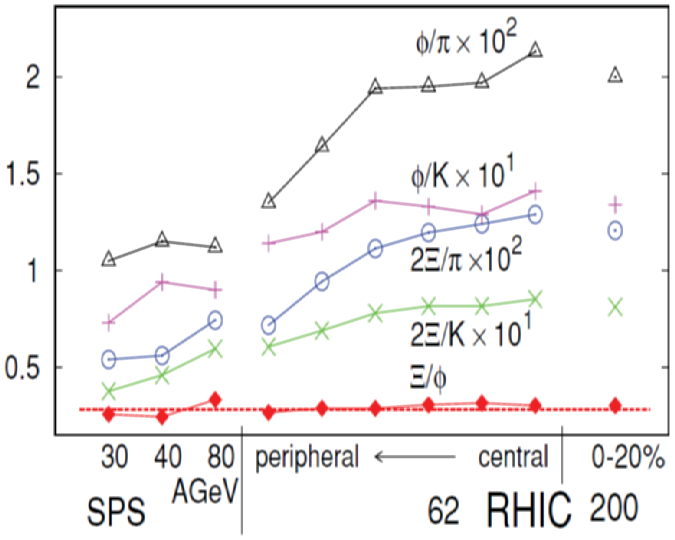}
}
\caption{Data points of $\Xi/\phi$ (bottom red diamonds) and a straight line at 0.281; compared to ratios of these particles with $\pi$ and K which are highly variable. To eliminate in baryon-rich domain dependence  on chemical potential a geometric mean of $\Xi$ with $\overline\Xi$ is shown. Adapted from Ref.\;\cite{Petran:2009dc}.\label{Canonical} }
\end{figure}

Another interesting result, combining RHIC and SPS data, is that the ratio $\Xi(ssq)/\phi(s\bar s)$ of those two different double strange particles is an energy independent constant (see \rf{Canonical}). This observation decisively resolves the discussion about canonical strangeness suppression, \ie\ the volume dependent suppression caused by overall net strangeness conservation. The data provide for a clear connection of multi-strange particle production to total strangeness content, but not to net strangeness content, which is zero for the $\phi$. This rules out canonical suppression as a viable explanation of the multiplicity dependence of strangeness enhancement seen in Figs.~\ref{WA85WA94AntiHFig}, \ref{AliceALL}, and \ref{Canonical}. Moreover, the universal value of the ratio is signaling that irrespective of how the fireball is formed there is no significant alteration in the final state of the yields of these double strange particles. For further analysis of the implications of these data we refer to Petran's work.\cite{Petran:2009dc}

Returning now to continue the LHC contribution to QGP physics, the decrease of the enhancement effect with increasing collision energy seen in \rf{AliceALL} signals that there is an increase in strange antihyperon production in the control $pp$ or $pA$ collisions. This effect has been nicely demonstrated by the ALICE-LHC collaboration, which recently published~\cite{ALICE:2017jyt} the enhancement over minimum bias $pp$ yields as a function of the charged hadron multiplicity $dN_{\rm ch}/d\eta$, combining  $AA$-results~\cite{ABELEV:2013zaa} with $pA$\cite{Adam:2015vsf} and new $pp$ results as shown in \rf{AliceNew} on the left. The $pp$ and $pA$ results merge into the $AA$ results when compared for the same multiplicity at central rapidity. We see a smooth increase with $dN_{\rm ch}/d\eta$, and in most cases a yield saturation at large hadron multiplicity indicating that QGP in internal thermal and chemical equilibrium is achieved as the volume of the fireball grows. 

\begin{figure}[htb]
\centerline{%
\includegraphics[width=\linewidth]{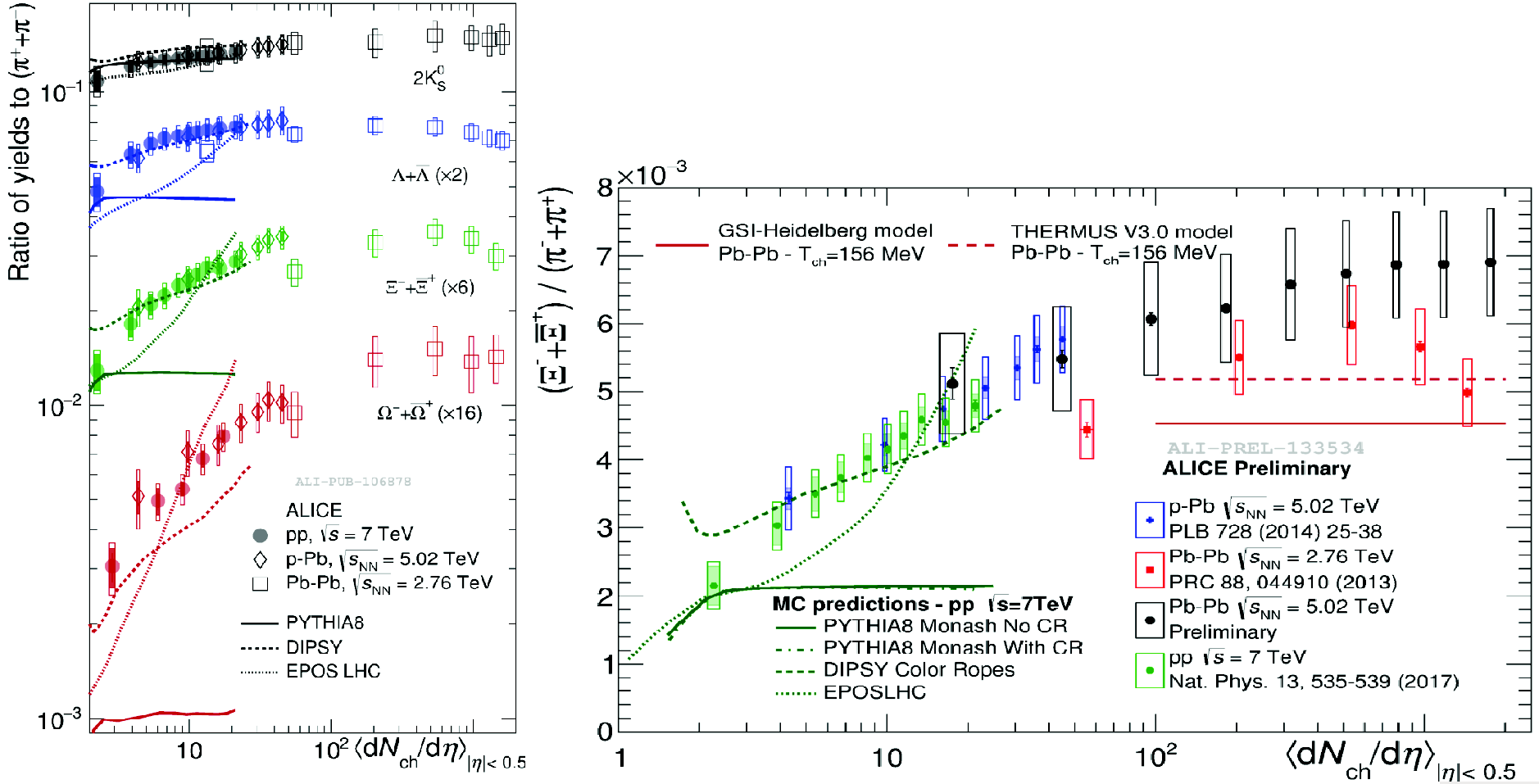}}
\caption{On left overview of results for strangeness signature of QGP from ALICE published Spring 2017~\cite{ALICE:2017jyt}. On right the results of July 2017 SQM meeting showing the need to revisit the old analysis of $\Xi$-yields.  QCD Monte-Carlo simulations are not able to reproduce the observed results (curves marked PYTHIA8, DIPSY, EPOS LHC).\label{AliceNew}}
\end{figure}

Such an equilibrated QGP fireball hadronizes into an out of chemical equilibrium hadron abundance, an important insight we discussed already in our 1986 review~\cite{Koch:1986ud}. The predominant modification of the fireball at hadronization for collisions in which  strangeness is chemically saturated in the QGP is the fireball hadronization volume.  A potential spoiler in the \rf{AliceNew} is the behavior of $\Xi$ in $AA$ which seems to be on the low end of the $pA$ results, an effect less than two standard deviations but somewhat disturbing the otherwise consistent pattern. Today this has attracted additional attention since the ALICE collaboration presented (preliminary) results at the highest LHC energy, $\sqrt{s_{\rm NN}}=5.02$ TeV, at the {\em SQM2017} conference. 

These new  $\sqrt{s_{\rm NN}}=5.02$ TeV  $\Xi$ results differ significantly from those at $\sqrt{s_{\rm NN}}=2.76$ TeV as shown in the right panel of \rf{AliceNew}. The remarkable, yet (for us) expected outcome is that the results at $\sqrt{s_{\rm NN}}=5.02$ are where we expected the data from $\sqrt{s_{\rm NN}}=2.76$ TeV to be, in agreement with the findings for other strange hadrons (results available, not shown in \rf{AliceNew}).  The ALICE Collaboration announced at SQM2017 the intent to review the analysis of the  $\Xi$ data from the 2012/13 $\sqrt{s_{\rm NN}}=2.76$ TeV runs --- the trends of $pp$ and $pA$ results suggest that the revised yields will agree with the new $\sqrt{s_{\rm NN}}=5.02$ TeV results.

Let us now assume that the corrected  $\sqrt{s_{\rm NN}}=2.76$ TeV antihyperons will indeed track the still preliminary $\sqrt{s_{\rm NN}}=5.02$ TeV data. This would demonstrate that the strangeness signature of QGP is driven by the global properties of thermal fireball, in particular its volume and/or life-time, and not by the collision system or the collision energy. This is so since the hadronization condition of QGP fireball is known to be universal across a wide range of collision energies and centralities~\cite{Letessier:2005qe,Petran:2013qla,Rafelski:2015cxa}. 

We conclude that CERN-SPS results have shown (a) the onset of deconfinement near $\sqrt{s_\mathrm{NN}}=7$\,GeV, and that the systematics of multi-strange particle production eliminates alternate enhancement mechanisms, as borne out \eg\ by the $\Xi/\phi$ ratio. The LHC data indicate that for the thousandfold higher LHC energy QGP can already be formed in high multiplicity $pp$ and $pA$ reactions. Across a wide range of accessible collision energies at SPS, RHIC, and LHC strangeness in the QGP saturates, and the fireball hadronizes in nearly identical fashion at universal physical conditions, with the final geometric size determining the total particle yields. The formation of the thermal QGP fireball depends on entropy content (hadron multiplicity) and not on how this state has been produced, the $pp$,  $pA$ and $AA$ collisions being equivalent.

\section{The BNL-RHIC contribution to the QGP discovery}\label{sec:BNL}
\subsection{Flow dynamics of QCD matter}\label{ssec:liquid}

The \lq\lq standard model\rq\rq\  of the dynamics of a relativistic heavy ion collision begins with a very brief period of kinetic equilibration -- most likely less than 1 fm/c. After that  the space-time evolution of QGP can be described by relativistic viscous hydrodynamics. Hydrodynamics is the effective theory of the transport of energy and momentum in matter on long distance and time scales. In order to be applicable to the description of QGP  created in \RHI\ collisions, which forms tiny, short-lived droplets of femtometer size, the hydrodynamic equations must be relativistic and include the effects of (shear) viscosity. The causal relativistic theory of viscous fluid has been worked out over the past decade. It is based on the framework of the M\"uller-Israel-Stewart formulation of second-order hydrodynamics, which includes relaxation effects for the dissipative part of the stress tensor. Schematically, the equations have the form $\partial_\mu T^{\mu\nu} = 0$ with
\begin{eqnarray}
T^{\mu\nu} & = & (\varepsilon +P) u^\mu u^\nu + \Pi^{\mu\nu} 
\\
\tau_\pi (d\Pi^{\mu\nu}/d\tau) + \Pi^{\mu\nu} & = & \eta (\partial^\mu u^\nu + \partial^\nu u^\mu - {\rm trace}) .
\end{eqnarray}

It turns out that the ratio of the shear viscosity $\eta$ to the entropy density $S$ jointly with the equation of state is the quantity that most directly controls the behavior of the fluid. The quantity $\eta/S$ is the relativistic generalization of the well known kinematic viscosity. Since in kinetic theory $\eta$ is proportional to the mean free path of particles in the fluid, which is inversely proportional to the transport cross section, unitarity limits how small $\eta$ can become under given conditions. An interesting consequence of this observation is that the quantity $\eta/S$ has an apparent lower bound of the order of 0.08 (in units of $\hbar$). The existence of such a bound was conjectured already three decades ago, but it was quantitatively derived only recently using the technique of holographic gravity duals. It is now believed that $\eta/S \ge (4\pi)^{-1}$ for most sensible quantum field theories~\cite{Kovtun:2004de} at $\mu_B \approx 0$. 

The experimental handle for the determination of $\eta/S$ is the azimuthal anisotropy of the flow of final-state particles in off-central heavy-ion collisions, where the nuclear overlap region is elongated in the direction perpendicular to the reaction plane. Hydrodynamics converts the anisotropy of the pressure gradient into a flow anisotropy, which sensitively depends on the value of $\eta/S$. The average geometric shape of the overlap region in symmetric nuclear collisions is dominated by the elliptic eccentricity, resulting an elliptic flow anisotropy characterized by the second Fourier coefficient $v_2$. Event-by-event fluctuations of the density distribution with \lq Standard\rq\ model of dn the overlap region generate higher Fourier coefficients for the initial geometry and final flow, encoded in higher Fourier coefficients $v_3$, $v_4$, etc. Their measurement is analogous to the mapping of the amplitudes of multipoles in the thermal fluctuations of the cosmic background radiation.

The precise results of such an analysis of event-by-event fluctuations of the flow distribution depends somewhat on the structure of the initial-state density fluctuations, especially their radial profile and spatial scale. The  study of this kind to date  starts from the fluctuations of the gluon distribution in the colliding nuclei, evolves them for a brief period using classical Yang-Mills equations, and then inserts the fluctuating energy density distribution into viscous hydrodynamics.~\cite{Gale:2012rq} The conclusion of this study is that the average value of $\eta/S$ (averaged over the thermal history of the expansion) in Au+Au collisions at the top RHIC energy is 0.12; whereas the value for Pb+Pb collisions at LHC is 0.20 (see \rf{BM:fig4}). While each of these values has systematic uncertainties of at least 50\%, the ratio of these two values is probably rather stable against changes in the assumptions for the initial state. 

\begin{figure}[!t]
\begin{center}
 \parbox{5in}{\includegraphics[width=5in]{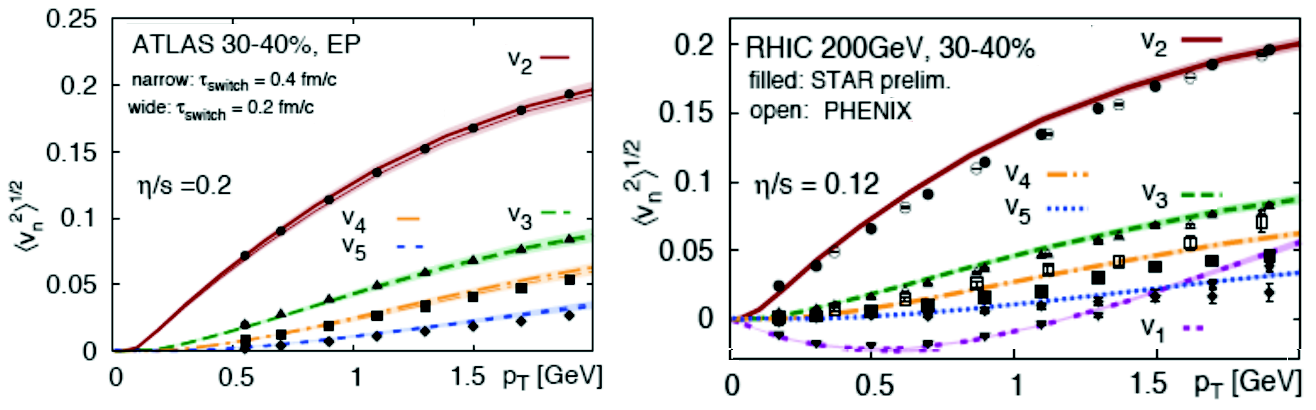}}
\caption{Fourier components of collective flow, $v_n(p_T)$, for Au+Au collisions at RHIC (left panel) and Pb+Pb collisions at LHC (right panel), in comparison with viscous hydrodynamics calculations. The deduced average value of the ratio $\eta/S$ is 60\% larger at LHC (0.20) than at RHIC (0.12).  (Adapted from Gale {\em et.al.}~\cite{Gale:2012rq})\label{BM:fig4}}
\end{center}
\end{figure}

One interpretation of this result is  that the average value of $\eta/S$ at the 10 times higher LHC energy is somewhat higher than at RHIC, indicating a significant temperature dependence of this quantity. In the frame of this explanation  the quark-gluon plasma at the lower temperature reached at RHIC is more strongly coupled and a more \lq\lq  perfect\rq\rq\  liquid, making this energy domain especially interesting for the study of this observable.~\cite{Bass:2017zyn} 

\subsection{Valence quark recombination}\label{ssec:reco}

If the term {\em quark-gluon plasma} is to truly apply to the hot QCD matter created in heavy ion collisions, it must contain excitations with the quantum numbers of quarks and gluons that are not confined into color singlet objects. As discussed in the context of the strangeness signature, one then expects that hadrons are formed by coalescence of valence quarks ~\cite{Biro:1983gh,Koch:1986ud,Rafelski:1987un} when the matter cools back down below $T_c$. This predicted the strongly enhanced production of hadrons containing multiple strange quarks as a characteristic signature of the formation of a quark-gluon plasma. This enhancement was clearly observed in Au+Au collisions at RHIC, as it had been before in $AA$ collisions at the CERN-SPS, but its value as a quark-gluon plasma signature had been questioned on the basis of the fact that it is already present at the lower collision energies and also occurs to a lesser degree in proton-nucleus collisions. Recent results from the beam energy scan at RHIC, which clearly indicate the presence of partonic collective behavior at the top SPS energy domain, and from p+Pb collisions at LHC and d+Au collisions at RHIC, which produced strong evidence for the presence of collective flow and showed a continuity of strangeness enhancement with final-state multiplicity, have all but eliminated these doubts. 

Support for  quark recombination idea came early in the RHIC physics program from particle identified spectra measured by PHENIX \cite{Adler:2003kg}. These showed an enhancement in the ratio of protons to pions in the transverse momentum range $p_T = 1-3$ GeV/c. This finding became known as the \lq\lq proton anomaly\rq\rq. The data also showed an apparent deviation from the mass hierarchy of the elliptic flow $v_2(p_T)$ of identified hadrons in the same momentum range \cite{Adler:2003kt}. Hydrodynamics predicts that heavier hadrons should exhibit a smaller flow anisotropy at the same momentum $p_T$, but PHENIX data showed that the $v_2$ of protons and antiprotons exceeds that of pions for $p_T > 2$ GeV/c.

The concept of valence quark recombination explained both experimental findings. If the collective transverse flow is carried by quarks and these quarks recombine at the moment of hadronization, then protons carrying three valence quarks receive a larger transverse momentum boost from the collective expansion than pions, which contain only two valence quarks. The same argument applies of course to all baryons and mesons. The application of the sudden recombination model relies on the insight that valence quarks coalescing into a hadron with a few GeV/c transverse momentum leave the quark-gluon plasma at nearly the speed of light and thus make a sudden transition from the dense matter into the surrounding vacuum.

\begin{figure}
\begin{center}
\includegraphics[width=4.in]{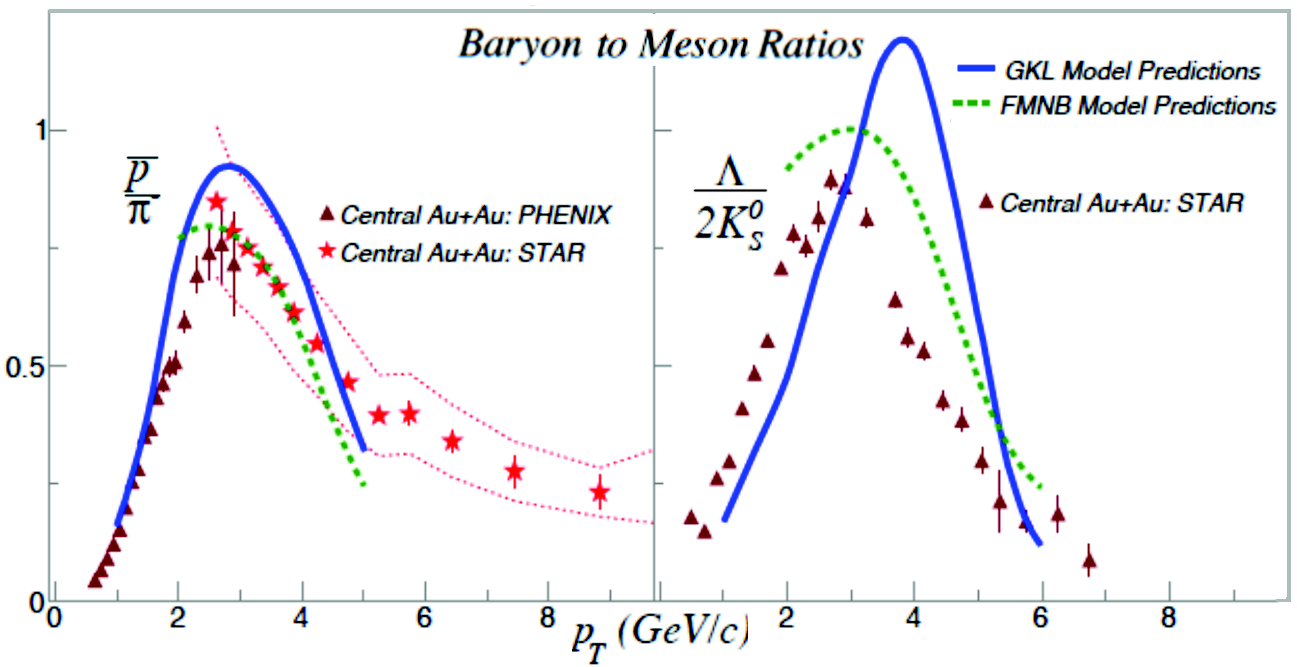}
\caption{Antiproton-to-pion ratio and hyperon-to-kaon ratio in Au+Au collisions as a function of transverse momentum in comparison with two quark recombination models. (Adapted from R.~J.~Fries \protect\cite{Fries:2011wz}.)\label{BM:fig5}}
\end{center}
\end{figure}

Theoretical considerations show that the mechanism of quark recombination from a thermal quark-gluon plasma \cite{Fries:2003vb,Greco:2003xt} with the transverse flow generated by the expansion at RHIC exceeds the contribution to hadron formation by parton fragmentation for transverse momenta $p_T < 3-4$ GeV/c, precisely the regime where the proton puzzle was observed. Using reasonable values for the expansion velocity at the moment when the cooling matter hadronizes led to quantitative predictions for the transverse momentum dependence of the $\overline{p}/\pi$ and $\Lambda/K_s^0$ ratios as well as the elliptic flow of protons and pions, which reproduced the essential features of the PHENIX and STAR data (see Fig.~\ref{BM:fig5}).

\begin{figure}[!b]
\begin{center}
\includegraphics[width=4.in]{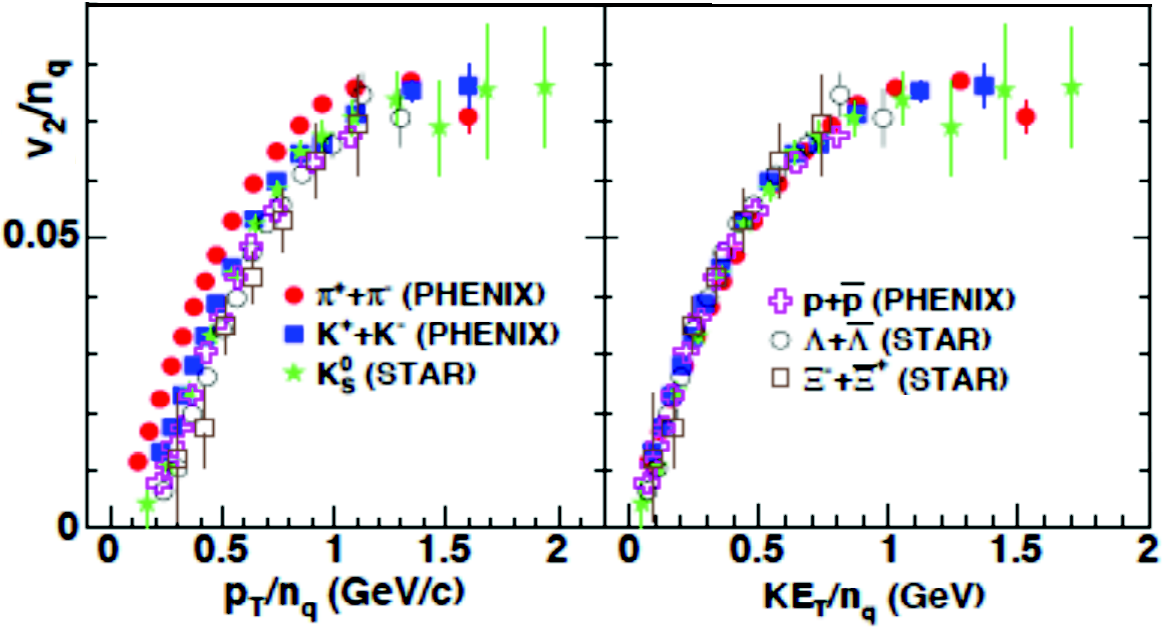}
\caption{Valence quark number scaled elliptic flow for various hadron species. Left panel: scaling by transverse momentum $p_T$; right panel: scaling by transverse mass $m_T$. (Adapted from Adare {\em et al.}\cite{Adare:2006ti}.)\label{BM:fig6}}
\end{center}
\end{figure}

A particular interesting relationship is obtained for the elliptic flow spectrum of different hadron species containing $n$ valence quarks \cite{Fries:2003kq}: $v_2(p_t) \approx n v_2^q(p_T/n)$, which relates the elliptic flow spectrum of mesons ($n=2$) to that of baryons ($n=3$). At low transverse momenta, where mass effects are not negligible, it has been suggested that the transverse momentum $p_T$ variable should be replaced by the transverse kinetic energy $m_T = \sqrt{p_T^2+m^2}$. With this heuristic substitution, the valence quark scaling of elliptic flow was found to hold over the entire range of available data \cite{Adare:2006ti} (see Fig.~\ref{BM:fig6}).

In a more differential way than the strangeness enhancement signature, the sudden recombination model for hadron emission from the nuclear fireball provided evidence for the formation of a new state that contains collectively flowing matter composed of independently moving quarks and antiquarks. In the summary of their original publication Fries {\em et al.}~\cite{Fries:2003vb} concluded: \lq\lq ...we propose a two component behavior of hadronic observables in heavy ion collisions at RHIC. These components include fragmentation of high-$p_T$ partons and recombination from a thermal parton distribution. $\ldots$ Our scenario requires the assumption of a thermalized partonic phase characterized by an exponential momentum spectrum. Such a phase may be appropriately called a quark-gluon plasma.\rq\rq

\subsection{Beam Energy Scan}\label{ssec:BES}

During the years 2010--2011 RHIC conducted a beam energy scan (BES) for Au+Au collisions that covered the energies $\sqrt{s_\mathrm{NN}} = 7.7, 11.5, 19.6, 27, 39~\mathrm{GeV}$. The energy provided an essential link between the results previously obtained at the CERN-SPS and the data measured at the full RHIC collision energy of $\sqrt{s_\mathrm{NN}} = 200~\mathrm{GeV}$. A rather complete compilation of the data for observables related to the chemical and kinetic freeze-out parameters of the medium can be found in Ref.~\cite{Adamczyk:2017iwn}. The BES confirmed the strong enhancement of strangeness production observed in the CERN-SPS experiments, including detailed features like Marek's \lq\lq horn\rq\rq, the peak in the $K^+/\pi^+$ ratio near $\sqrt{s_\mathrm{NN}} = 7~\mathrm{GeV}$. The strangeness flavor was found to be chemically equilibrated in central collisions over the entire energy range, and the main difference between different energies can be attributed to a strong variation in the baryon chemical potential from $\mu_B \approx 100~\mathrm{MeV}$ at the high-energy end to $\mu_B \approx 400~\mathrm{MeV}$ at the low-energy end of the BES. In summary, the BES established the continuity of chemical and thermal bulk properties of the medium from the CERN-SPS to the BNL-RHIC energy range and provided compelling evidence that a quark-gluon plasma is temporarily created across this range of collision energies.

\subsection{Jet Quenching}\label{ssec:jets}

Energetic partons, the precursors of later emerging hadronic jets, lose energy while traversing the quark gluon plasma either by elastic collisions with the medium constituents or by gluon radiation. At high energies, radiation should dominate; collisional energy loss is expected to be important for intermediate energy partons and for heavy quarks. Each mechanism is encoded in a transport coefficient, $\hat{e}$ for collisional energy loss and $\hat{q}$ for radiative energy loss:~\cite{Majumder:2010qh}
\begin{equation}
(dE/dx)_{\rm coll} = - C_2 \hat{e} , \qquad  (dE/dx)_{\rm rad} = - C_2 \hat{q} L ,
\end{equation}
where $L$ denotes the path length traversed in matter and $C_2$ is the quadratic Casimir of the fast parton. The value of $\hat{q}$ is given by the transverse momentum broadening of a fast light parton per unit path length.

The evolution of a jet in the medium, shown schematically in \rf{JETcone}, is characterized by several scales: The initial virtuality $Q_{\rm in}$ associated with the hard scattering process; the transverse scale at which the medium appears opaque, also called the saturation scale $Q_s$; and the transverse geometric extension of the jet, $r_\perp$. Those components of the jet, for which $r_\perp > Q_s^{-1}$, will be strongly modified by the medium. This means that the core of the jet will remain rather inert, except for an overall energy attenuation of the primary parton, but strong modifications are expected at larger angles and soft components of the jet. These features of jet modification  can be encoded in a transport equation for the accompanying gluon radiation:~\cite{Qin:2010mn}
\begin{equation}
\frac{d}{dt}f(\omega,k_\perp^2,t) = \hat{e} \frac{\partial f}{\partial\omega} + \frac{\hat{q}}{2} \frac{\partial f}{\partial k_\perp^2}
+ \frac{dN_{\rm rad}}{d\omega dk_\perp^2 dt} ,
\end{equation}
where the last term denotes the gluon radiation induced by the medium.
 
\begin{figure}[htb]
\begin{center}
\includegraphics[width=4in]{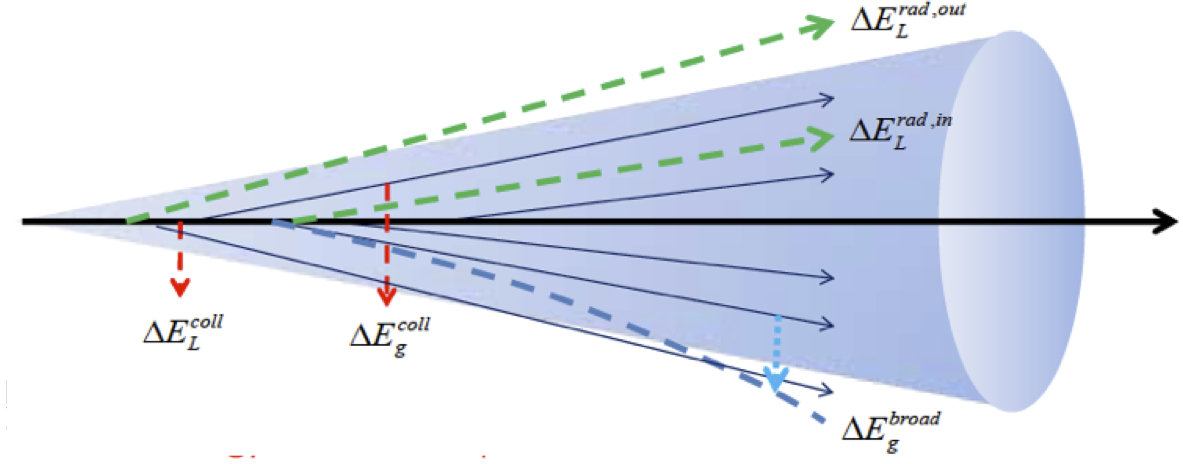}
\caption{Schematic representation of matter induced processes that contribute to jet quenching, \ie\ the loss of energy out of the jet cone.\label{JETcone}}
\end{center}
\end{figure}

\begin{figure}[!b]
\begin{center}
\includegraphics[width=3in]{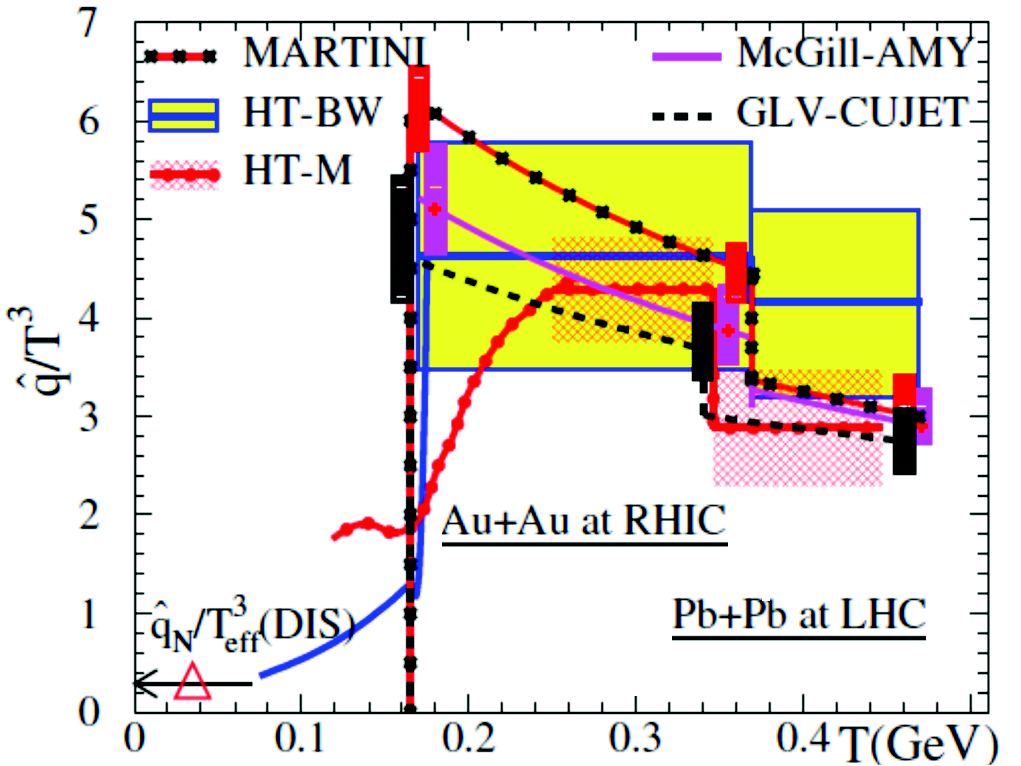}
\caption{Temperature scaled jet quenching parameter $\hat{q}$ deduced by the JET Collaboration from measurements of inclusive hadron suppression at RHIC and LHC. (Adapted from Burke {\em et al.} \cite{Burke:2013yra}.)\label{BM:fig10}}
\end{center}
\end{figure}

The \lq\lq  jet quenching parameter\rq\rq\  $\hat{q}$ can be determined by analyzing the suppression of leading hadrons in A+A collisions, compared with the scaled p+p data, usually given by a suppression factor $R_{AA}$, which is of the order of 0.2 for hadrons of transverse momenta in the range of 10 MeV/c in Au+Au at RHIC and Pb+Pb at LHC. A systematic analysis of available data from RHIC and LHC was published by the JET Collaboration \cite{Burke:2013yra} (see Fig.~\ref{BM:fig10}). It suggests that the temperature averaged value of $\hat{q}$ grows slightly less than linear with the matter density between RHIC and LHC. This confirms the notion that the quark-gluon plasma formed at higher temperatures is somewhat less strongly coupled.

Using the values of $\hat{q}$ and $\hat{e}$ determined by comparison with $R_{AA}$ data, one can also explain the strong increase of the di-jet asymmetry observed in central Pb+Pb collisions at LHC. This gives confidence that the basic mechanisms of jet modification and parton energy loss are reasonably well understood. The phenomenology of jet quenching at the LHC, exemplified by the CMS data from Pb+Pb collisions~\cite{Chatrchyan:2011sx}  shown in \rf{BM:fig7}, agrees qualitatively well with the expectation that modifications are concentrated at large cone angles and soft momentum fractions within the jet.  

\begin{figure}[!t]
\begin{center}
\includegraphics[width=4.5in]{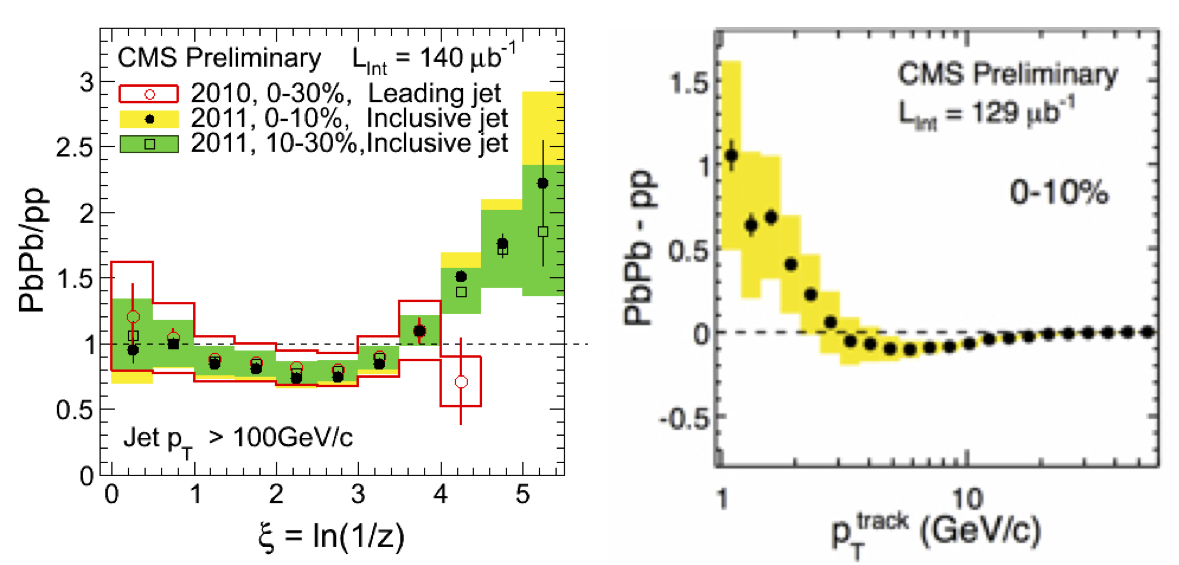}
\caption{Jet modifications observed in Pb+Pb collisions at LHC by the CMS collaboration  (Adapted from Chatrchyan {\em et al.}~\cite{Chatrchyan:2011sx}, CMS collaboration.)\label{BM:fig7}}
\end{center}
\end{figure}

\subsection{Quarkonium Melting}\label{ssec:QQ}

Bound states of heavy quarks, especially quarkonia ($J/\psi$, $\psi'$, the $\Upsilon$ states), are sensitive to the distance at which the color force is screened in the quark-gluon plasma. Several mechanisms contribute to nuclear modification of the quarkonium yield as is illustrated in \rf{BM:fig8}. At sufficiently high temperatures the screening length becomes shorter than the size of the quarkonium radius and the $Q\overline{Q}$ bound state \lq\lq  melts\rq\rq\ . Since the radii of the quarkonium states vary widely -- from approximately 0.1fm for the $\Upsilon$ ground state to almost 1 fm for $\psi'$ -- the sequential melting of these states could enable at least a semi-quantitative determination of the color screening length.  

\begin{figure}[!b]
\begin{center}
\includegraphics[width=4.2in]{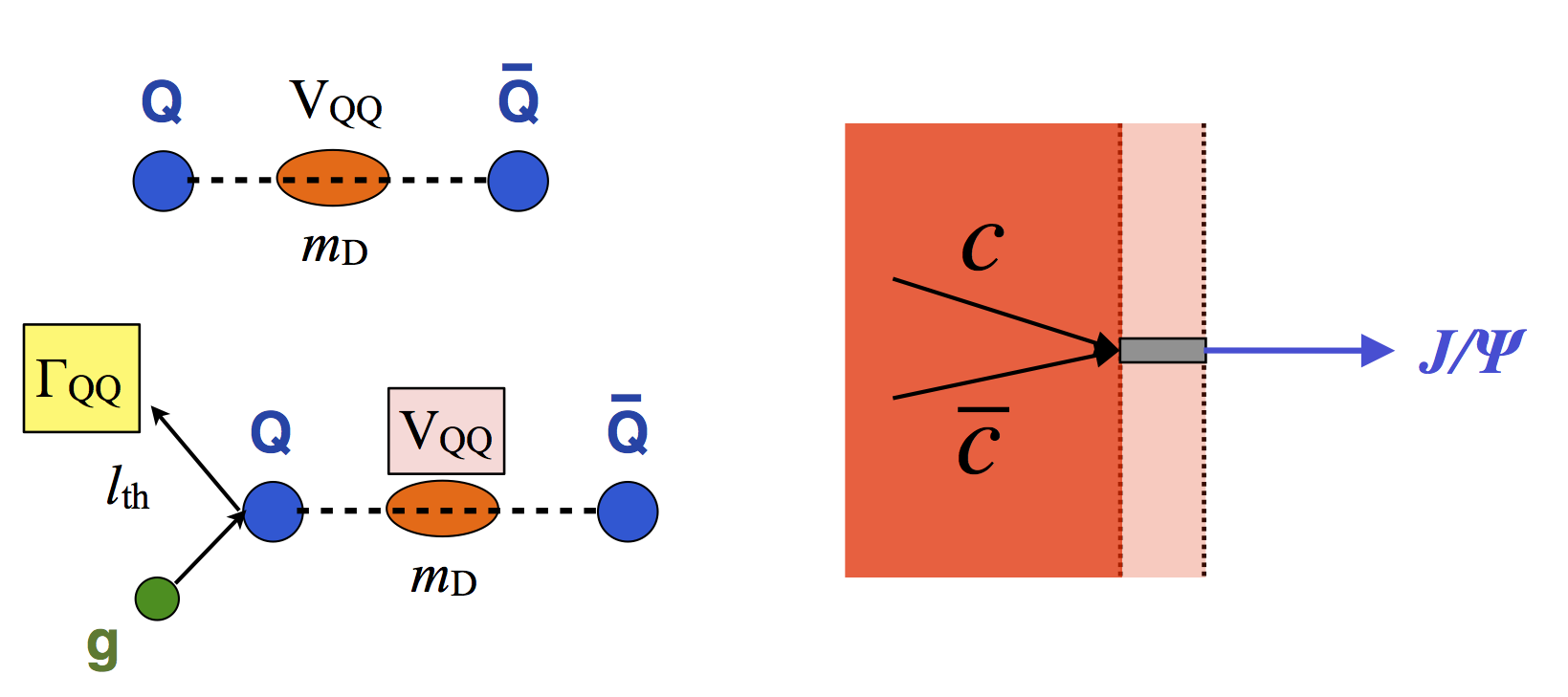}
\caption{Mechanisms contributing to matter induced changes in the yield of quarkonia: Color screening (upper left); ionization by thermal gluons (lower left); and recombination (right). 
\label{BM:fig8}}
\end{center}
\end{figure}

The static screening length, which is relevant for heavy quarks, can be calculated within the context of  lattice QCD. However, it has become well understood in recent years that static color screening is only part of the picture of quarkonium melting, and that quarkonium yields can not only be suppressed by the action of the medium, but also enhanced by recombination, if the density of heavy quarks and antiquarks is large enough. An important loss mechanism is ionization by absorption of thermal gluons. This mechanism gains in importance as the binding energy of a quarkonium state is lowered by color screening. The absorption channel can be included in the dynamical evolution of the amplitude as an imaginary part of the potential with a corresponding noise term ensuring ultimate approach to the equilibrium distribution:
\begin{equation}
i\hbar\frac{\partial}{\partial t} \Psi_{Q\overline{Q}} = \left[ \frac{p_Q^2 + p_{\overline{Q}}^2}{2M_Q} + V_{Q\overline{Q}} - \frac{i}{2}\Gamma_{Q\overline{Q}} + \xi_{Q\overline{Q}} \right] \Psi_{Q\overline{Q}} .
\end{equation}

Recombination of a heavy $Q\overline{Q}$-pair can occur at or near hadronization, similar to the sudden recombination mechanism that is thought to be responsible for the valence quark scaling of the identified particle elliptic flow. The yield of quarkonia formed in this manner grows quadratically with the heavy quark yield. Recombination of charm quark pairs into $J/\psi$ and $\psi'$ is thus expected to be much more frequent at LHC than at RHIC. This expectation is borne out by a comparison of the centrality dependence of $J/\psi$ suppression observed by PHENIX in Au+Au at RHIC and by ALICE in Pb+Pb at LHC (see \rf{BM:fig9}).
 
\begin{figure}[!b]
\begin{center}
 \parbox{2.5in}{\includegraphics[width=2.4in]{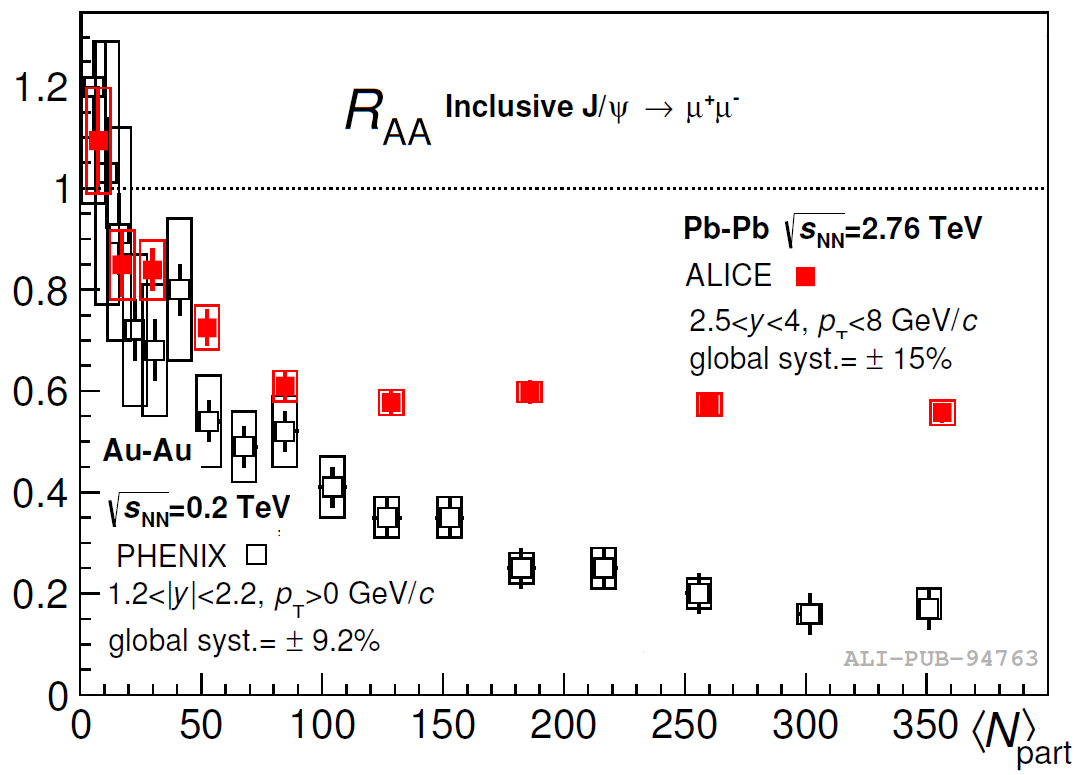}}
 \parbox{2.2in}{\includegraphics[width=2.1in]{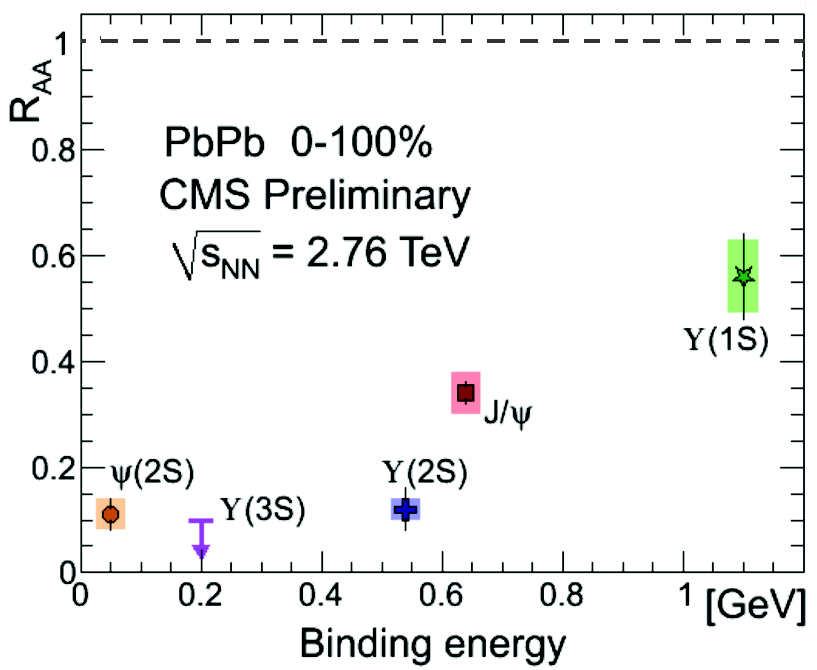}}
\caption{Left panel: Nuclear suppression factor $R_{AA}$ for $J/\psi$ as function of collision centrality in Au+Au at RHIC (red) and Pb+Pb at LHC (blue). Right panel: $R_{AA}$ in Pb+Pb at LHC for different quarkonium states. The surprisingly small amount of suppression for $J/\psi$ at LHC is evidence for recombination.\label{BM:fig9} (Adapted from Andronic~\cite{Andronic:2013awa} [ALICE] and Calderon de la Barca Sanchez~\cite{Sanchez:2013lpa} [CMS].)}
\end{center}
\end{figure}

The LHC data show {\em less} suppression in central collisions than the RHIC data, although the significantly hotter matter proceed at LHC energy must surely be more effective in melting the $J/\psi$ state. Whether it is possible to measure enough observables in order to not only disentangle the action of these different mechanisms, but also determine the color screening length, will need to be seen. On the positive side, the theory of quarkonium transport in hot QCD matter has now reached a state of sophistication where this seems possible.

In summary of this  section we note that the RHIC-BNL  contributed several additional and convincing experimental observables which further evolved comparing to the contemporary LHC results (which we did not discuss beyond strangeness). Without further discussion let us say that there is reason for hope that details of the initial state structure can be separated from viscous effects, and both can be separately extracted from the data. Jet physics opens new avenues of probing the quark-gluon plasma at different scales. The quarkonium data from the LHC suggest that recombination dominates in central Pb+Pb collisions for the $c\overline{c}$ states.

\section{Summary}

Since we have offered sub-summaries at the end of each section, we can be brief here: We have described in some detail the experimental developments that followed on the publication of our review of the strangeness signature of the QGP~\cite{Koch:1986ud} in 1986. While over the past 30 years much has been learned in terms of experimental data accumulation and theoretical analysis, the basic insights described in our review have withstood the test of time. No other viable interpretation of all soft hadron production data measured in \RHI\ collisions exists than the formation of a thermally and chemically equilibrated QGP fireball that explosively disintegrates preserving entropy, strangeness and baryon number content established at the boundary between QGP and hadron gas. The fleeting presence of the QGP is most clearly witnesses by the large overabundance of strange antibaryons. 

Our chronology of the CERN-SPS strangeness research shows that the key experimental results were available as early as 1992 and were several times confirmed and published before CERN finally announced the QGP discovery in February 2000. We have made an attempt to explain why this announcement was less broadly accepted than probably would have been the case if it had been primarily based on the observation of the predicted and quantitatively confirmed strangeness enhancement. As it was presented in 2000, many doubts remained, and a whole new set of experiments at BNL-RHIC was required to establish a broad consensus with respect to the discovery of a new state of matter on the basis of additional phenomena not accessible at the SPS energies.

We did not dwell on the numerous attempts made over the past 30 years to question the usefulness of strangeness as signature of QGP, as none of the related arguments has withstood the test of time.  However, as an example of such proposals we discussed the fact that the $\Xi/\phi$ ratio is constant over the whole range of SPS and RHIC energies while the degree of strangeness enhancement varies substantially, discrediting the family of models called \lq canonical enhancement\rq. 

With new results for strange baryon and antibaryon enhancement over a wide range of system sizes now emerging from experiments at LHC, it becomes possible to explore in detail how the strangeness flavor is chemically equilibrated as function of QGP lifetime and size. As charm becomes an abundant flavor in the LHC energy domain, flavor probes of the QGP are further expanding their reach as primary bulk signals and probes of QGP formation. While the basic picture is now well established, much still waits to be learned.


\end{document}